\documentclass[11pt]{article}

\usepackage{color}
\usepackage{latexsym}
\usepackage{amssymb}
\usepackage{graphicx}
\usepackage{caption}

\usepackage{amsmath, amsfonts,amssymb,theorem,euscript,
array,enumerate,amsfonts,mathrsfs}

\newtheorem{Theorem}{Theorem}[part]
\newtheorem{Definition}{Definition}[part]
\newtheorem{Proposition}{Proposition}[part]

\newtheorem{Lemma}{Lemma}[part]

\newtheorem{Remark}{Remark}[part]

\renewcommand{\theDefinition}{\thesection.\arabic{Definition}}
\renewcommand{\theTheorem}{\thesection.\arabic{Theorem}}
\renewcommand{\theProposition}{\thesection.\arabic{Proposition}}

\renewcommand{\theLemma}{\thesection.\arabic{Lemma}}
\renewcommand{\theRemark}{\thesection.\arabic{Remark}}

\renewcommand{\theequation}{\thesection.\arabic{equation}}

\def\qed{\hbox{\hskip 6pt\vrule width6pt height7pt
depth1pt  \hskip1pt}\bigskip}

\def \Frac{\displaystyle\frac}

\def \I{\mathbb{I}}
\def \N{\mathbb{N}}
\def \R{\mathbb{R}}

\def \E{\mathbb{E}}
\def \F{\mathbb{F}}

\def \P{\mathbb{P}}

\def \Ac{{\cal A}}
\def \Bc{{\cal B}}
\def \Cc{{\cal C}}

\def \Ec{{\cal E}}
\def \Fc{{\cal F}}

\def \Ic{{\cal I}}

\def \Lc{{\cal L}}
\def \Pc{{\cal P}}

 \def \Nc{{\cal N}}
\def \Sc{{\cal S}}

\def \Uc{{\cal U}}
\def \Vc{{\cal V}}
\def \Wc{{\cal W}}

\def \Vc{{\cal V}}

\def \eps{\varepsilon}

\def \ep{\hbox{ }\hfill$\Box$}

\def\Dx#1{\Frac{\partial #1}{\partial x}}

\def\Dy#1{\Frac{\partial #1}{\partial y}}

\def\Dyy#1{\Frac{\partial^2 #1}{\partial y^2}}

\def\reff#1{{\rm(\ref{#1})}}

\def\beqs{\begin{eqnarray*}}
\def\enqs{\end{eqnarray*}}
\def\beq{\begin{eqnarray}}
\def\enq{\end{eqnarray}}

\title{Investment/consumption problem in illiquid markets with regime-switching}
\author{Paul Gassiat$^{1)}$, Fausto Gozzi$^{2)}$, Huy\^en
Pham$^{1),3)}$}

\begin{document}
\maketitle {\noindent
\small
\begin{tabular}{llll}
$^{1)}$ & Laboratoire de Probabilit\'es et              &  $^{2)}$ & Dipartimento di Scienze Economiche   \\
        & Mod\`eles Al\'eatoires,  CNRS, UMR 7599              &          & ed Aziendali - Facolt\`a di Economia,\\
        &  Universit\'e Paris 7 Diderot,     &          & Universit\`a LUISS Guido Carli, \\
        &      pgassiat, pham  at math.jussieu.fr                    &  & fgozzi at luiss.it  \\
$^{3)}$ &          CREST-ENSAE                    &           &   \\
        &             and Institut Universitaire de France                                  &          &    \\[1ex]
\end{tabular} }

\begin{abstract}
We consider an illiquid financial market with different regimes modeled by a conti\-nuous-time finite-state Markov chain. The investor can trade a stock only at the discrete arrival times of a Cox process with intensity depending on the market regime. Moreover, the risky asset price is subject to liquidity shocks, which
change its rate of return and volatility, and induce jumps on its dynamics. In this setting, we study the problem of an economic agent optimizing her expected utility
from consumption under a non-bankruptcy constraint.  By using the dynamic programming method, we provide the characterization of the value function of this stochastic control pro\-blem in terms of the unique viscosity solution to a system of   integro-partial differential equations. We next focus on the popular case of CRRA utility functions, for which we can prove smoothness $C^2$ results for the value function. As an important byproduct, this allows us to get the existence of optimal investment/consumption strategies characterized in feedback forms. We analyze a  convergent numerical scheme for the resolution to our stochastic control problem, and  we illustrate finally with some numerical experiments the effects of liquidity regimes in the investor's optimal decision.
\end{abstract}

\vspace{5mm}

\noindent {\bf Key words~:}  Optimal consumption, liquidity effects, regime-switching models,  viscosity solutions, integro-differential system.

\vspace{5mm}


\vspace{2mm}

\noindent {\bf MSC Classification (2000)~:} 49K22, 49L25,  60J75, 91B28, 93E20.

\newpage


\section{Introduction}

\setcounter{equation}{0} \setcounter{Assumption}{0}
\setcounter{Theorem}{0} \setcounter{Proposition}{0}
\setcounter{Corollary}{0} \setcounter{Lemma}{0}
\setcounter{Definition}{0} \setcounter{Remark}{0}

A classical assumption in the theory of optimal portfolio/consumption choice as in Merton \cite{mer71}  is that assets are continuously tradable by agents.  This is not always realistic in practice, and
illiquid markets provide a prime example. Indeed, an important aspect of market liquidity is the time restriction on assets trading: investors cannot buy and sell them immediately, and have to wait some time before being able to unwind a position in some financial assets.  In the past years,  there was a significant strand of literature addressing these liquidity constraints.  In \cite{rogzan02}, \cite{mat06},
the price process is observed continuously  but the trades succeed only at the jump times of a Poisson process. Recently, the papers \cite{phatan08}, \cite{cregozphatan11}, \cite{gasphasir10}  relax the continuous-time
price observation by considering that asset is observed only at the random trading times.  In all these cited papers, the intensity of  trading times is constant or deterministic. However, the market liquidity is also affected by
long-term macroeconomic conditions, for example by financial crisis or political turmoil, and so the level of trading activity measured by its intensity should vary randomly over time.
Moreover,  liquidity breakdowns would typically induce drops on the stock price in addition to changes in its rate of return and volatility.

In this paper, we investigate the effects of such liquidity features on the optimal portfolio choice.  We model the index of  market liquidity as an observable continuous-time Markov chain with finite-state regimes, which is consistent with some
cyclicality observed in financial markets. 
The modelisation of financial stock prices by regime-switching processes was originally proposed and justified in \cite{ham89}, and since then this approach has been extensively pursued in the financial litterature, see e.g. \cite{angbek02}, \cite{sotcad09} and the references therein.

The economic agent can trade only at the discrete arrival times of a Cox process with intensity depending on the market regimes.  Moreover, the risky asset price is subject to liquidity shocks, which switch its rate of return and
volatility, while inducing jumps on its dynamics.  In this hybrid jump-diffusion setting with regime switching, we study the optimal investment/consumption problem over an infinite horizon under a nonbankruptcy state constraint. 
We first prove carefully that dynamic programming principle (DPP) holds in our framework.  Due to the state constraints in two dimensions, 
we have to slightly weaken the standard continuity assumption, see Remark \ref{rm:proofcontinuity}. Then, using DPP, 
we characterize the value function of this stochastic control problem as the unique constrained viscosity solution to a system of integro-partial differential equations. In the particular case of CRRA utility function, we can go beyond the viscosity properties, and prove $C^2$ regularity results for the value function in the interior of the domain. As a consequence, we show  the existence of optimal strategies expressed in feedback form in terms of the derivatives of the value function. Due to the presence of state constraints, the value function is not smooth at the boundary, and so the verification theorem cannot be proved with the classical arguments of Dynkin's formula.   To overcome this technical problem, we use  an ad hoc approximation procedure (see Proposition \ref{propVerif}). 
We also provide a convergent numerical scheme for solving the system of equations characterizing our control problem, and we illustrate with some numerical results the effect of liquidity regimes in the agent's optimal investment/consumption. We also measure the impact of continuous time observation with respect to a discrete time observation of the stock prices.
Our paper contributes and extends the existing literature in several ways. First, we extend the papers \cite{rogzan02} and \cite{mat06} by considering stochastic intensity trading times and regime switching in the asset prices.
For a  two-state Markov chain modulating the market liquidity, and in the limiting case where the intensity in one regime goes to infinity, while the other one goes to zero, we recover the setup of \cite{diekrasei09} and \cite{ludmin10}
where an investor can trade continuously in the perfectly liquid regime but faces a threat of trading interruptions  during a period of market freeze.
On the other hand,  regime switching models in optimal investment problems was already used in \cite{zari02}, \cite{sotcad09} or \cite{pirzha11}  for continuous-time trading.

The rest of the paper is structured as follows. Section 2 describes  our continuous-time  market model with regime-switching liquidity, and formulates the optimization problem for the investor. In Section 3 we state some useful properties of the value function of our stochastic control problem.  Section 4 is devoted to the analytic characterization of the value function as the unique viscosity solution to the dynamic programming equation.
The special case of CRRA utility functions is studied  in Section 5:  we show smoothness results for the value functions, and obtain the  existence of  optimal strategies via a verification theorem.  Some numerical illustrations
complete this last section. Finally two appendices are devoted to the proof of two technical results: the dynamic programming principle, and the existence and uniqueness of viscosity solutions.

\section{A market model with regime-switching liquidity}

\setcounter{equation}{0} \setcounter{Assumption}{0}
\setcounter{Theorem}{0} \setcounter{Proposition}{0}
\setcounter{Corollary}{0} \setcounter{Lemma}{0}
\setcounter{Definition}{0} \setcounter{Remark}{0}

Let us fix a probability space $(\Omega,\Fc,\P)$ equipped with a filtration $\F$ $=$ $(\Fc_t)_{t\geq 0}$ satisfying the usual conditions.   It is assumed that all random variables and stochastic processes are defined on the stochastic basis $(\Omega,\Fc,\F,\P)$.

Let $I$ be a  continuous-time Markov chain  valued in the finite state space $\I_d$ $=$ $\{1,\ldots,d\}$, with intensity matrix $Q$ $=$
$(q_{ij})$.  For $i$ $\neq$ $j$ in $\I_d$,  we can associate to the jump process $I$, a Poisson process $N^{ij}$ with intensity rate $q_{ij}$ $\geq$ $0$,
such that a switch from  state $i$ to $j$ corresponds to a jump of $N^{ij}$ when $I$ is in state $i$. 
We interpret the process $I$ as a proxy for market liquidity  with states (or regimes) representing the level of liquidity activity,
in the sense that  the intensity of trading times varies with the regime value. This is modeled through a Cox process $(N_t)_{t\geq 0}$ with intensity
$(\lambda_{I_t})_{t\geq 0}$, where $\lambda_i > 0$ for each $i \in \I_d$  . For example, if $\lambda_i$ $<$ $\lambda_j$, this means that trading times occur more often
in regime $j$ than in regime $i$.  The increasing sequence of jump times $(\tau_n)_{n\geq 0}$, $\tau_0$ $=$ $0$,  associated to the counting process $N$ represents the random times  when  an investor can trade a risky asset of price process $S$. Note that under these assumptions the jumps of $I$ and $N$ are a.s. disjoint. 


In the liquidity regime $I_t$ $=$ $i$, the stock price follows the dynamics
\beqs
dS_t &=& S_t (b_i dt + \sigma_i dW_t),
\enqs
where $W$ is a standard Brownian motion independent of $(I,N)$, and $b_i$ $\in$ $\R$, $\sigma_i$ $\geq$ $0$, for $i$ $\in$ $\I_d$. Moreover,
at the times of transition from $I_{t^-}$ $=$ $i$ to $I_t$ $=$ $j$, the stock
changes as follows:
$$S_t=S_{t -}(1-\gamma_{ij})$$
for a given $\gamma_{ij} \in (-\infty, 1)$, so the stock price remains strictly positive, and we may have a relative loss (if $\gamma_{ij}$ $>$ $0$), or gain (if $\gamma_{ij}$ $\leq$ $0$).
Typically, there is a drop of the stock price after a liquidity breakdown, i.e.  $\gamma_{ij}$ $>$ $0$ for $\lambda_j$ $<$ $\lambda_i$. Overall, the risky asset is governed by a regime-switching jump-diffusion model:
\beq \label{dynS}
dS_t &=& S_{t^-} \Big(b_{I_{t^-}}  dt + \sigma_{I_{t^-}}   dW_t -  \gamma_{_{I_{t^-},I_t}} dN^{I_{t^-},I_t}_t \Big).
\enq

{\it Portfolio dynamics under liquidity constraint.}   We consider an agent investing and consuming
in this regime-switching market. We denote by $(Y_t)$  the total
amount  invested in the stock, and by $(c_t)$ the  consumption rate per unit of  time, which is a nonnegative adapted process.
 Since the number of shares $Y_t/S_t$ in the stock held by the investor has to be kept constant between two trading dates $\tau_n$ and $\tau_{n+1}$,
then between such trading times, the process $Y$ follows the dynamics:
\beqs
dY_t &=& Y_{t^-} \frac{dS_t}{S_{t^-}}, \;\;\;
\tau_n \leq t < \tau_{n+1}, \; n \geq 0,
\enqs
The trading strategy is represented by a predictable process $(\zeta_t)$ such that at a trading time $t$ $=$ $\tau_{n+1}$,  the rebalancing on the number of shares  induces a jump $\zeta_t$ in  the amount  invested in the stock :
\beqs
\Delta Y_{t} &=&
\zeta_t.
\enqs
Overall, the c\`adl\`ag process $Y$  is governed by the hybrid controlled jump-diffusion process
\beq \label{dynY}
dY_t &=&  Y_{t^-} \Big(b_{I_{t^-}}  dt + \sigma_{I_{t^-}}   dW_t -  \gamma_{_{I_{t^-},I_t}} dN^{I_{t^-},I_t}_t \Big)  + \zeta_t dN_t.
\enq
Assuming for simplicity a constant savings account (see Remark \ref{rem:interestrate}), i.e. zero interest rate, the amount $(X_t)$ invested in cash then follows
\beq \label{dynX}
dX_t &=& -c_t dt - \zeta_t dN_t.
\enq
The total wealth is defined at any time $t$ $\geq$ $0$, by $R_t$ $=$ $X_t$ $+$ $Y_t$, and we shall require the non-bankruptcy constraint at any trading time:
\beq \label{eq:nonbankruptcy}
R_{\tau_{n}} & \geq & 0, \;\;\; a.s. \;\;\; \forall n \geq 0.
\enq
Actually since the asset price may become arbitrarily large or small between two trading dates, this non-bankruptcy constraint means  a no-short sale constraint on both the stock and savings account, as showed by the following Lemma. 

\begin{Lemma} \label{pr:constraints}
The nonbankruptcy constraint (\ref{eq:nonbankruptcy})  is  formulated equivalently in the no-short sale constraint:
\beq \label{eq:nonbankruptcyequivstate}
 X_t \ge 0, & \mbox{ and } & Y_t \ge 0, \;\;\; \forall t \geq 0.
\enq
This is also written equivalently in terms of the controls as:
\beq
    & & - Y_{t^-} \; \leq \; \zeta_t \; \leq \; X_{t^-}, \;\;\;t \geq 0, \label{eq:nonbankruptcyequivcontrol1} \\
\int_t^{\tau_{n+1}} c_s ds & \leq & X_t , \;\;\; \tau_n \leq t < \tau_{n+1}, \; n \geq 0.  \label{eq:nonbankruptcyequivcontrol2}
    \enq
\end{Lemma}
{\bf Proof.} By writing by induction the wealth at any trading time as
\beqs
R_{\tau_{n+1}} &=& R_{\tau_n} + Y_{\tau_n}\left(\frac{S_{\tau_{n+1}}}{S_{\tau_n}}-1\right) - \int_{\tau_n}^{\tau_{n+1}} c_t dt, \;\;\; n \geq 0,
\enqs
and since (conditionally on $\Fc_{\tau_n}$) the stock price $S_{\tau_{n+1}}$ has support in $(0,\infty)$, we see that the nonbankruptcy condition $R_{\tau_{n+1}}$ $\geq$ $0$
is equivalent to a no-short sale constraint:
\beq \label{noshort1}
0 \; \leq \; Y_{\tau_n} \; \leq \; R_{\tau_n}, \;\;\; n \geq 0,
\enq
together with the condition on the nonnegative consumption rate
\beq \label{admic1}
\int_{\tau_n}^{\tau_{n+1}} c_t dt & \leq &  R_{\tau_n} - Y_{\tau_n} = X_{\tau_n}, \;\;\; n \geq 0.
\enq
Since $Y_{\tau_n} = Y_{\tau_n-}+\zeta_{\tau_n}$,
and since $R_{\tau_n} = R_{\tau_n-}$ a.s.,
the no-short sale constraint \reff{noshort1} means
equivalently that \reff{eq:nonbankruptcyequivcontrol1} is satisfied for $t=\tau_n$. Since $\zeta$ is predictable, this is equivalent to \reff{eq:nonbankruptcyequivcontrol1} being satisfied $d\P \otimes dt$ almost everywhere. Indeed, letting $H_t= \mathbf{1}_{\left\{\zeta_t<-Y_{t-} \mbox{ or } \zeta_t >  X_{t^-} \right\}}$, $H$ is predictable, so that $\forall t \geq 0$, 
$0 = \E\left[\sum_{\tau_n \leq t} H_{\tau_n}\right] = \E\left[ \int_0^t H_s \lambda_{I_s}ds\right] $, and we deduce that $H_t$ $=$ $0$ $d\P\otimes dt$ a.e. since $\lambda_{I_t}$ $>$ $0$.

  Moreover, since $X_t$ $=$ $X_{\tau_n} - \int_{\tau_n}^t c_s ds$ for $\tau_n \leq t < \tau_{n+1}$,
the condition \reff{admic1} is equivalent to \reff{eq:nonbankruptcyequivcontrol2}. By rewriting the conditions \reff{noshort1}-\reff{admic1} as
\beqs
Y_{\tau_n} \geq 0, \;\;\;  X_{\tau_n} \geq 0, \;\;\;  X_{(\tau_{n+1})-} \geq 0, \;\;\; \forall n\geq 0,
\enqs
and observing that for $\tau_n \leq t < \tau_{n+1}$,
\beqs
Y_t \; =\;  \frac{S_t}{S_{\tau_n}}Y_{\tau_n}, & &  X_{\tau_n} \; \geq \; X_t \;  \geq X_{(\tau_{n+1})-},
\enqs
we see that they are equivalent to \reff{eq:nonbankruptcyequivstate}.
\ep

\begin{Remark} \label{remRZ}
{\rm Under the nonbankruptcy (or no-short sale constraint), the wealth $(R_t)_{t\geq 0}$ is nonnegative, and follows the dynamics:
\beq
\label{dynR}
dR_t &=&   R_{t^-} Z_{t^-} \Big(b_{I_{t^-}}  dt + \sigma_{I_{t^-}}   dW_t -  \gamma_{_{I_{t^-},I_t}} dN^{I_{t^-},I_t}_t \Big) - c_t dt,
\enq
where $Z_t:= \frac{Y_t}{R_t} $ valued in  $[0,1]$ is the proportion of  wealth invested in the risky asset; and evolves according to the dynamics:
\beq
\label{dynZ}
dZ_t &=& Z_{t^-}(1-Z_{t^-}) \Big[  \big(b_{I_{t^-}} - Z_{t^-} \sigma_{I_{t^-}}^2\big) dt  + \sigma_{I_{t^-}} dW_t  -
\frac{\gamma_{_{I_{t^-},I_t}} }{1- Z_{t^-} \gamma_{_{I_{t^-},I_t}} }  dN^{I_{t^-},I_t}_t \Big]  \nonumber \\
& & \;\;\;\;\;\;\;\;\;\;\;  + \;  \frac{\zeta_t}{R_{t-}} dN_t \; + \; Z_{t-} \frac{c_t}{R_{t-}}dt,
\enq
for $t$ $<$ $\tau$ $=$ $\inf\{ t \geq 0: R_t = 0\}$.\hfill\qed
}
\end{Remark}

Given an initial  state  $(i,x,y)$ $\in$ $\I_d\times\R_+\times\R_+$, we shall denote by  $\Ac_i(x,y)$ the set of
investment/consumption control process $(\zeta,c)$ such that the corresponding  process $(X,Y)$ solution to
\reff{dynY}-\reff{dynX} with a liquidity regime $I$, and starting from  $(I_{0^-},X_{0^-},Y_{0^-})$ $=$ $(i,x,y)$, satisfy the non-bankruptcy constraint
\reff{eq:nonbankruptcyequivstate} (or equivalently \reff{eq:nonbankruptcyequivcontrol1}-\reff{eq:nonbankruptcyequivcontrol2}).

\vspace{2mm}

{\it Optimal investment/consumption problem.}  The preferences of the agent are described by a utility function $U$ which is increasing, concave, $C^1$ on $(0,\infty)$ with $U(0)=0$, and satisfies the usual Inada conditions: $U'(0)$ $=$ $\infty$, $U'(\infty)$ $=$ $0$.  We assume the following growth condition on $U$ : there exist some positive constant $K$, and $p$ $\in$ $(0,1)$ s.t.
\beq \label{growthU}
U(x) &\leq& K x^p, \;\;\; x \geq 0.
\enq
We denote by $\tilde U$ the convex conjugate of $U$, defined from $\R$ into $[0,\infty]$ by:
\beqs
\tilde U(\ell) &=& \sup_{x\geq 0} [U(x) - x \ell],
\enqs
which satisfies under \reff{growthU} the dual growth condition on $\R_+$:
\beq \label{dualgrowth}
\tilde U(\ell) & \leq & \tilde K \ell^{-\tilde p}, \;\;\;\forall \ell \geq 0, \;\;\; \mbox{ with } \; \tilde p \; = \; \frac{p}{1-p} \; > \; 0,
\enq
for some positive constant $\tilde K$.

The agent's objective is to maximize over portfolio/consumption strategies in the above illiquid market model the expected utility from consumption rate  over an infinite horizon.  We then consider, for each $i \in \I_d$, the value function
\beq
v_i(x,y) &=& \sup_{(\zeta,c)\in\Ac_i(x,y)} \E\left[ \int_0^\infty e^{-\rho t} U(c_t) dt  \right], \;\;\;
(x,y) \in \R_+^2, \label{defVi}
\enq
where $\rho$  is a discount factor. We also introduce, for $i \in \I_d$, the function
\beq \label{defhatvi}
\hat{v}_i(r) &=&  \sup_{x \in [0,r]} v_i(x,r-x), \;\;\;
 r \geq 0,
\enq
which represents the maximal utility performance that the agent can achieve starting from an initial nonnegative wealth $r$ and from the regime $i$.
More generally, for  any locally bounded function $w_i$ on $\R_+^2$, we associate  the function $\hat w_i$ defined on $\R_+$ by:  $\hat w_i(r)$ $=$ $\sup_{x\in [0,r]} w_i(x,y)$, so that:
\beqs
\hat w_i(x+y) &=& \sup_{ e \in [-y,x]} w_i(x-e,y+e), \;\;\; (x,y) \in \R_+^2.
\enqs
In the sequel, we shall often identify a $d$-tuple  function $(w_i)_{i\in\I_d}$ defined on $\R_+^2$ with the function $w$ defined on  $\R_+^2\times\I_d$ by $w(x,y,i)$ $=$ $w_i(x,y)$.

In this paper, we focus on the analytic characterization of the value functions $v_i$ (and so $\hat v_i$),
$i$ $\in$ $\I_d$, and on their numerical approximation.

\begin{Remark} \label{rem:interestrate}
{\rm For simplicity we have assumed zero interest rate for the riskless asset. The case of constant $r \neq 0$ can actually be reduced to this case, at the cost of allowing time-dependent utility of consumption. This can be seen from the identity
\beqs
\E \int_0^\infty e^{- \rho s} U(c_s) ds &=& \E \int_0^\infty e^{- \bar \rho s} \bar U(s, \bar c_s) ds,
\enqs
where $\bar \rho = \rho - p r$, $\bar c_s = e^{-r s} c_s$ and $\bar U(s,\bar c) = e^{-prs}U(e^{rs} \bar c)$. Note that $\bar U(s,\cdot)$ still satisfies \reff{growthU}, and in the special case of power utility $U(c) = c^p$, one actually has $\bar U(s,\cdot)=U$.
}\hfill\qed
\end{Remark}
\section{Some properties of the value function}

\setcounter{equation}{0} \setcounter{Assumption}{0}
\setcounter{Theorem}{0} \setcounter{Proposition}{0}
\setcounter{Corollary}{0} \setcounter{Lemma}{0}
\setcounter{Definition}{0} \setcounter{Remark}{0}

We state some preliminary properties of the value functions that will be used in the next section for the PDE characterization.
We first need to check that the value functions are well-defined and finite. Let us consider for any $p$ $>$ $0$, the positive constant:
\beqs \label{AssRho}
k(p) &:=&  \max_{i \in \I_d,z \in  [0,1]} \Big[ p b_i z - \frac{\sigma_i^2}{2} p (1-p) z^2 + \sum_{j \neq i} q_{ij}((1 - z \gamma_{ij})^p - 1)  \Big] \; < \; \infty.
\enqs
We then have the following lemma.

\begin{Lemma} \label{lemGrowthR}
Fix some initial conditions $(i,x,y)$ $\in$ $\I_d\times\R_+\times\R_+$, and some $p$ $>$ $0$. Then:
\begin{itemize}
  \item [{\bf (1)}]
  For any admissible control $(\zeta,c)$ $\in$ $\Ac_i(x,y)$ associated with wealth process $R$, the process
$(e^{-k(p)t}R_t^p)_{t\geq 0}$ is a supermartingale.  So, in particular, for $\rho$ $>$ $k(p)$,
\beq \label{limRp}
\lim_{t\rightarrow\infty} e^{-\rho t} \E[R_t^p] &=& 0.
\enq
  \item [\bf (2)] For fixed $T$ $\in$ $(0,\infty)$, the family $(R_{T\wedge\tau}^p)_{\tau,\zeta,c}$ is uniformly integrable, when $\tau$ ranges over all stopping times, and $(\zeta,c)$ runs over $\Ac_i(x,y)$.
\end{itemize}
\end{Lemma}
{\bf Proof.} {\bf (1)} By  It\^o's formula and \reff{dynR}, we have
\beqs
d(e^{-k(p)t} R_t^p) &=& - k(p) e^{-k(p)t} R_t^p dt + e^{-k(p)t}  d(R_t^p)   \\
&=& e^{-k(p)t} \Big[- k(p) R_t^p + p R_{t-}^{p-1} \left( - c_t + b_{I_{t-}} R_{t-} Z_{t-} \right)+ \frac{p(p-1)}{2} R_{t-}^{p-2}
\big(\sigma_{I_{t-}}R_{t-} Z_{t-}\big)^2  \\
& &    \;\;\;\;\;\;\;\;\;\;\;\;   + \;  \sum_{j \neq I_{t-}} q_{I_{t-},j}(R_{t-}^p(1 - \gamma_{I_{t-}j} Z_{t-})^p - R_{t-}^p)  \Big] dt  + dM_t,
\enqs
where $M$ is a local martingale. Now, by definition of $k(p)$, we have
\beqs
p R_{t-}^{p-1} \left( - c_t + b_{I_{t-}} R_{t-} Z_{t-} \right)+ \frac{p(p-1)}{2} R_{t-}^{p-2} \big(\sigma_{I_{t-}} R_{t-} Z_{t-}\big)^2  & & \\
+ \sum_{j \neq I_{t-}} q_{I_{t-},j}(R_{t-}^p(1 - \gamma_{I_{t-}j} Z_{t-})^p - R_{t-}^p)  &\leq& - p c_t R_{t-}^{p-1} + k(p) R_{t-}^p \\
&\leq& k(p) R_{t-}^p.
\enqs
Since $R$ has countable jumps, $R_t = R_{t-}$, $d\P \otimes dt$ a.e., and so the drift term in $d(e^{-k(p)t} R_t^p)$ is nonpositive.
Hence $(e^{-k(p)t}R_t^p)_{t\geq 0}$ is a local supermartingale, and since it is nonnegative,  it is a true supermartingale by Fatou's lemma.
In particular, we have
\beq \label{limRp2}
0 \leq e^{-\rho t} \E[R_t^p]\leq e^{-(\rho-k(p))t} (x+y)^p
\enq
which shows  \reff{limRp}.

\noindent {\bf (2)}  For any $q$ $>$ $1$, we get by the supermartingale property of the process $(e^{-k(pq)t} R_t^{pq})_{t\geq 0}$ and the optional sampling theorem:
\beqs
\E\big[\big(R_{T\wedge\tau}^p\big)^q\big] & \leq & e^{k(pq)T} (x+y)^{pq}\; < \;  \infty, \;\;\; \forall (\zeta,c) \in \Ac_i(x,y), \; \tau \mbox{ stopping time },
\enqs
which proves the required uniform integrability.
\ep

\vspace{2mm}

The next proposition states a comparison result, and, as a byproduct, a growth condition for the value function.

\begin{Proposition} \label{compaclassi}
\begin{itemize}
  \item[]
  \item[{\bf (1)}] Let $w$ $=$ $(w_i)_{i\in\I_d}$ be a $d$-tuple of nonnegative functions on $\R_+^2$, twice differentiable on $\R_+^2\setminus\{(0,0)\}$ such that
\beq
\rho w_i -  b_i y \Dy{w_i}  -  \frac{1}{2} \sigma_i^2 y^2 \Dyy{w_i}  - \sum_{j\neq i} q_{ij} [ w_j(x,y(1-\gamma_{ij})) - w_i(x,y)] & & \nonumber \\
\;\;\; - \; \lambda_i   [ \hat w_i(x+y) - w_i(x,y)]  - \tilde U\left(\Dx{w_i}\right) & \geq & 0, \label{surw}
\enq
for all $i$ $\in$ $\I_d$, $(x,y)$ $\in$ $\R_+^2\setminus\{(0,0)\}$.   Then, for all $i$ $\in$ $\I_d$,
$v_i$ $\leq$ $w_i$, on  $\R_+^2$.

  \item[{\bf (2)}]
  Under \reff{growthU},  suppose that $\rho$ $>$ $k(p)$. Then, there exists some positive constant $C$ s.t.
\beq \label{growthvi}
v_i(x,y) & \leq & C(x+y)^p, \;\;\; \forall (i,x,y) \in \I_d\times\R_+^2.
\enq
\end{itemize}
\end{Proposition}
{\bf Proof.} {\bf (1)} First notice that for $(x,y)$ $=$ $(0,0)$, the only admissible control in $\Ac_i(x,y)$ is the zero control
$\zeta$ $=$ $0$, $c$ $=$ $0$, so that $v_i(0,0)$ $=$ $0$. Now, fix $(x,y)$ $\in$ $\R_+^2\setminus\{(0,0)\}$, $i$ $\in$ $\I_d$, and
consider an arbitrary admissible control $(\zeta,c)$ $\in$ $\Ac_i(x,y)$. By It\^o's formula to $e^{-\rho t}w(X_t,Y_t,I_t)$, we get:
\beq
d[e^{-\rho t} w(X_t,Y_t,I_t)] &=& e^{-\rho t} \Big[ - \rho w - c_t  \Dx{w} +  b_{I_{t^-}} Y_{t^-} \Dy{w} + \frac{1}{2}\sigma_{I_{t^-}}^2 Y_{t^-}^2 \Dyy{w}   \nonumber \\
& &  \;\; + \;  \sum_{j\neq I_{t^-} } q_{_{I_{t^-}j}} [ w(X_{t^-},Y_{t^-}(1-\gamma_{_{I_{t^-}j}}),j) - w(X_{t^-},Y_{t^-},I_{t^-})]  \nonumber \\
& & \;\; + \;  \lambda_{_{I_{t^-}}} \big[ w(X_{t^-} - \zeta_t,Y_{t^-}+\zeta_t,I_{t^-}) - w(X_{t^-},Y_{t^-},I_{t^-}) \big] \Big] dt \nonumber \\
& & \;\;\; + \;  e^{-\rho t} \sigma_{I_{t^-}}^2 Y_{t^-} \Dy{w}(X_{t^-},Y_{t^-},I_{t^-}) dW_t \nonumber \\
& &  +    e^{-\rho t} \sum_{j\neq I_{t^-} }  [ w(X_{t^-},Y_{t^-}(1-\gamma_{_{I_{t^-}j}}),j) - w(X_{t^-},Y_{t^-},I_{t^-})]
\big(  dN^{I_{t^-}j} - q_{_{I_{t^-}j}}  dt \big) \nonumber \\
& & +  e^{-\rho t}  \big[ w(X_{t^-} - \zeta_t,Y_{t^-}+\zeta_t,I_{t^-}) - w(X_{t^-},Y_{t^-},I_{t^-}) \big] \Big]  \big(dN_t - \lambda_{_{I_{t^-}}}  dt \big).  \label{ito}
\enq
Denote by $\tau$ $=$  $\inf\{t\geq 0: (X_t,Y_t) = (0,0)\}$, and consider the
sequence of bounded stopping times $\tau_n$ $=$ $\inf\{ t \geq 0: X_t+Y_t\geq n \mbox{ or } X_t+Y_t\leq 1/n\}$ $\wedge$ $n$,
$n$ $\geq$ $1$.  Then, $\tau_n$ $\nearrow$
$\tau$ a.s. when $n$ goes to infinity, and $c_t$ $=$ $0$, $X_t$ $=$ $Y_t$ $=$ $0$ for $t$ $\geq$ $\tau$, and so
\beq \label{reltau}
\E\Big[ \int_0^\infty e^{-\rho t} U(c_t) dt \Big] &=& \E \Big[ \int_0^\tau e^{-\rho t} U(c_t) dt \Big].
\enq
From It\^o's formula \reff{ito} between time $t$ $=$ $0$ and $t$ $=$ $\tau_n$, and observing that the integrands of the local martingale parts are bounded for $t$ $\leq$ $\tau_n$, we obtain after taking expectation:
\beqs
w(x,y,i) &=& \E \Big[  e^{-\rho \tau_n } w(X_{\tau_n},Y_{\tau_n},I_{\tau_n})  \\
& & \;\;  + \int_0^{\tau_n} e^{-\rho t} \Big(  \rho w +   c_t  \Dx{w}  -   b_{I_{t^-}} Y_{t^-} \Dy{w} -  \frac{1}{2}\sigma_{I_{t^-}}^2 Y_{t^-}^2 \Dyy{w}  \\
& & \;\;\;\;\;\;\;\;\;   - \sum_{j\neq I_{t^-} } q_{_{I_{t^-}j}} [ w(X_{t^-},Y_{t^-}(1-\gamma_{_{I_{t^-}j}}),j) - w(X_{t^-},Y_{t^-},I_{t^-})]  \\
& &  \;\;\;\;\;\;\;\;\;   - \lambda_{_{I_{t^-}}} \big[ w(X_{t^-} - \zeta_t,Y_{t^-}+\zeta_t,I_{t^-}) - w(X_{t^-},Y_{t^-},I_{t^-}) \big] \Big) dt \Big]  \\
& \geq &  \E \Big[  e^{-\rho \tau_n } w(X_{\tau_n},Y_{\tau_n},I_{\tau_n})  + \int_0^{\tau_n} e^{-\rho t} U(c_t) dt \Big] \; \geq \;
\E \Big[  \int_0^{\tau_n} e^{-\rho t} U(c_t) dt \Big],
\enqs
where we used  \reff{surw}, and the nonnegativity of $w$.  By sending $n$ to infinity with  Fatou's lemma,  and \reff{reltau}, we obtain the required
inequality: $w_i$ $\geq$ $v_i$ since $(c,\zeta)$ are arbitrary.

\vspace{1mm}

\noindent {\bf (2)} Consider the function $w_i(x,y)$ $=$ $C(x+y)^p$. Then, for $(x,y)$ $\in$ $\R_+^2\setminus\{(0,0)\}$, and denoting by $z$ $=$
$y/(x+y)$ $\in$ $[0,1]$, a straightforward calculation shows that
\beq
& & \rho w_i -  b_i y \Dy{w_i}  -  \frac{1}{2} \sigma_i^2 y^2 \Dyy{w_i}  - \sum_{j\neq i} q_{ij} [ w_j(x,y(1-\gamma_{ij})) - w_i(x,y)]  \nonumber \\
& & \;\;\; - \; \lambda_i    [ \hat w_i(x+y) - w_i(x,y)]  - \tilde U(\Dx{w_i})   \nonumber \\
&=&  C(x+y)^p \Big[ \rho - p b_i z +  \frac{\sigma_i^2}{2} p (1-p) z^2 -  \sum_{j \neq i} q_{ij}((1 - z \gamma_{ij})^p - 1) \Big] -
\tilde{U}((x+y)^{p-1} p C) \nonumber \\
&\geq& (x+y)^p \Big(C (\rho - k(p)) - \tilde{K} (pC)^{- \frac{p}{1-p}} \Big)  \label{inegw}
\enq
by \reff{dualgrowth}. Hence, for $\rho$ $>$ $k(p)$, and for $C$ sufficiently large, the r.h.s. of \reff{inegw} is nonnegative, and we conclude
by using the comparison result in assertion 1).
\ep

\vspace{2mm}

In the sequel, we shall assume the standing condition that $\rho$ $>$ $k(p)$ so that the value functions are well-defined and satisfy the growth condition \reff{growthvi}. We now prove continuity properties of the value functions.

\begin{Proposition} \label{propv}
The value functions $v_i$, $i$ $\in$ $\I_d$,  are concave, nondecreasing in both variables, and continuous on
$\R_+^2$.  This implies also that $\hat{v}_i$, $i$ $\in$ $\I_d$,  are nondecreasing, concave and continuous on $\R_+$. Moreover, we have the boundary conditions for $v_i$,  $i$ $\in$ $\I_d$, on $\{0\}\times\R_+$:
\beq \label{vix=0}
v_i(0,y) &=& \left\{  \begin{array}{cc}
			   0, & \mbox{ if } \; y = 0 \\
			 \E \Big[ e^{-\rho\tau_1} \hat v_{_{I_{\tau_1}^i}}\big( y \frac{S_{\tau_1}}{S_0}\big) \Big], &  \mbox{ if } \;  y > 0.
			 \end{array}
			 \right.
\enq
Here $I^i$ denotes the continuous-time Markov chain  $I$ starting from $i$ at time $0$.
\end{Proposition}

\noindent{\bf Proof.} Fix some $(x,y,i)$ $\in$ $\R_+^2\times\I_d$, $\delta_1 \geq 0$, $\delta_2$ $\geq$ $0$,
and take an admissible control $(\zeta,c)$ $\in$ $\Ac_i(x,y)$. Denote by $R$ and $R'$ the wealth processes associated to $(\zeta,c)$,  starting from initial state $(x,y,i)$ and $(x+\delta_1,y+\delta_2,i)$.  We thus have $R' = R+\delta_1+\delta_2 S/S_0$.
This implies that  $(\zeta,c)$ is also an admissible control for $(x+\delta_1,y+\delta_2,i)$, which shows clearly the nondecreasing
monotonicity of $v_i$ in $x$ and $y$, and thus also the nondecreasing monotonicity of  $\hat v_i$ by its very definition.

The concavity of $v_i$ in $(x,y)$ follows from the linearity of the admissibility constraints in $X,Y,\zeta,c$, and the concavity of $U$.
This also implies the concavity of $\hat{v_i}(r)$ by its definition.
%
%


Since $v_i$ is concave, it is  continuous on the interior of its domain $\R_+^2$. From \reff{growthvi}, and since $v_i$ is nonnegative, we see that
$v_i$ is continuous on $(x_0,y_0)$ $=$ $(0,0)$ with $v_i(0,0)$ $=$ $0$. Then, $\hat v_i$ is continuous on $\R_+$ with $\hat v_i(0)$ $=$ $0$.
It remains to prove the continuity of $v_i$ at  $(x_0,y_0)$ when  $x_0=0$ or $y_0=0$. We shall rely on the following implication of the dynamic programming principle
\beq
v_i(x,y) &=&  \sup_{c\in\Cc(x)} \E\Big[ \int_0^{\tau_1} e^{-\rho t} U(c_t) dt  + e^{-\rho\tau_1} \hat v_{_{I_{\tau_1}^i}}(R_{\tau_1}) \Big]  \label{dynvi1} \\
&=& \sup_{c\in\Cc(x)} \E\Big[ \int_0^{\tau_1} e^{-\rho t} U(c_t) dt
+ e^{-\rho\tau_1} \hat v_{_{I_{\tau_1}^i}} \big( x - \int_0^{\tau_1}c_t dt  + y \frac{S_{\tau_1}}{S_0} \big) \Big], \; \forall (x,y) \in \R_+^2, \nonumber
\enq
where $\Cc(x)$ denotes the set of nonnegative adapted processes $(c_t)$ s.t. $\int_0^{\tau_1} c_t dt$ $\leq$ $x$ a.s.

\noindent (i) We first consider the case $x_0$ $=$ $0$  (and $y_0$ $>$ $0$).

\noindent  In this case, the constraint on consumption $c$ in  $\Cc(x_0)$  means  that $c_t =0$, $t \leq \tau_1$, so that \reff{dynvi1} implies
\reff{vix=0}.  Now, since $v_i$ is nondecreasing in $x$, we have: $v_i(x,y) \geq v_i(0,y)$. Moreover, by concavity and thus continuity
of $v_i(0,.)$, we have: $\lim_{y \to y_0} v_i(0,y)=v_i(0,y_0)$.  This implies that $\liminf_{(x,y) \rightarrow (0,y_0)} v_i(x,y) \geq v_i(0,y_0)$.
The proof of the converse inequality requires more technical arguments. For any $x,y \geq 0$, we have:
\beq \label{cont1}
v_i(x,y) &=& \sup_{c \in \Cc(x)} \E \Big[ \int_0^{\tau_1} e^{-\rho s} U(c_s) ds + e^{-\rho \tau_1}\hat{v}_{_{I_{\tau_1}^i}}\big(x - \int_0^{\tau_1} c_s ds + y\frac{S_{\tau_1}}{S_0}\big) \Big] \nonumber \\
&\leq& \sup_{c \in \Cc(x)} \E \Big[ \int_0^{\tau_1} e^{-\rho s} U(c_s) ds \Big] \;
+ \;  \E \Big[ e^{-\rho \tau_1}\hat{v}_{I_{\tau_1}}\big(x + y\frac{S_{\tau_1}}{S_0}\big) \Big] \nonumber \\
&=:& E_1(x) + E_2(x,y).
\enq
Now, by Jensen's inequality, and since $U$ is concave,  we have:
\beqs
\int_0^\infty U\left(c_s \mathbf{1}_{\left\{s \leq \tau_1 \right\}}\right) \rho e^{-\rho s} ds &\leq& U\left(\int_0^\infty c_s \mathbf{1}_{\left\{s \leq \tau_1 \right\}} \rho e^{-\rho s} ds \right),
\enqs
and thus:
\beq \label{jensen}
\int_0^{\tau_1} e^{-\rho s} U(c_s) ds &\leq&
\frac{U(\rho x)}{\rho}, \;\;\;\;\;  a.s.  \;\;\; \forall c \in \Cc(x),
\enq
by using the fact that $\int_0^{\tau_1} c_t dt$ $\leq$ $x$ a.s.  By continuity of $U$ in $0$ with $U(0)$ $=$ $0$, this shows that $E_1(x)$ converges  to zero when $x$ goes to $x_0$ $=$ $0$. Next, by continuity of  $\hat{v}_i$, we have:
$\hat{v}_{_{I_{\tau_1}^i}}\big(x + y\frac{S_{\tau_1}}{S_0}\big)$  $\rightarrow$
$\hat{v}_{_{I_{\tau_1}^i}}\big(y_0\frac{S_{\tau_1}}{S_0}\big)$ a.s. when $(x,y) \rightarrow (0,y_0)$. Let us check  that this convergence is dominated. Indeed from \reff{growthvi}, there is some positive constant $C$ s.t.
\beqs
\hat{v}_{_{I_{\tau_1}^i}}\big(x + y \frac{S_{\tau_1}}{S_0}\big) &\leq& C \big(x+y\frac{S_{\tau_1}}{S_0}\big)^p \; \leq \;
C(x+y)^p \Big(1 \vee \Big(\frac{S_{\tau_1}}{S_0}\Big)^p\Big).
\enqs
Moreover,
\beqs
\E\Big[e^{- \rho \tau_1} \Big(\frac{S_{\tau_1}}{S_0}\Big)^p \Big| I,W \Big] &=&
\int_0^{\infty} \lambda_{I_t} e^{- \int_0^t \lambda_{I_s}} e^{-\rho t}\Big(\frac{S_t}{S_0}\Big)^p dt
\; \leq \;  \max_{i\in\I_d} \lambda_i  \; \int_0^{\infty}  e^{-\rho t} \left(\frac{S_t}{S_0}\right)^p dt,
\enqs
and so
\beqs
\E\Big[e^{-\rho \tau_1}\Big(\frac{S_{\tau_1}}{S_0}\Big)^p \Big] &\leq&
\max_{i\in\I_d} \lambda_i  \; \int_0^{\infty} \E\Big[e^{-\rho t}\Big(\frac{S_t}{S_0}\Big)^p\Big] dt \\
& \leq &  \max_{i\in\I_d} \lambda_i  \; \int_0^{\infty} e^{-(\rho -k(p))t} dt \; < \; \infty,
\enqs
where we used in the second inequality the supermartingale property in Lemma \ref{lemGrowthR} (and, more precisely, equation (\ref{limRp2})) for  $x=0,y=1, c \equiv \zeta \equiv 0$.
One can then apply the dominated convergence theorem to $E_2(x,y)$, to deduce that  $E_2(x,y)$ converges to
$\E\Big[e^{-\rho\tau_1}\hat{v}_{_{I_{\tau_1}^i}}\big(y_0\frac{S_{\tau_1}}{S_0}\big)\Big]$  when $(x,y) \rightarrow (0,y_0)$. This, together with
\reff{vix=0}, \reff{cont1}, proves that  $\limsup_{(x,y) \rightarrow (0,y_0)} v_i(x,y)$ $\leq$ $v_i(0,y_0)$, and thus the continuity of $v_i$ at $(0,y_0)$.

\vspace{2mm}

\noindent (ii) We consider the case  $y_0$ $=$ $0$ (and $x_0$ $>$ $0$).

\noindent Similarly, as in the first case, from the nondecreasing and continuity properties of $v_i(.,0)$, we have:
$\liminf_{(x,y) \rightarrow (x_0,0)} v_i(x,y) \geq v_i(x_0,0)$. Conversely, for any $x$ $\geq$ $0$, and $c \in \Cc(x)$, let us consider the stopping time  $\tau_c$ $=$ $\inf \big\{ t \in \geq 0: \int_0^t c_s ds = x_0 \big\}$. Then, the nonnegative adapted process $c'$ defined by:
$c'_t$ $=$ $c_t \mathbf{1}_{\big\{t \leq \tau_c \wedge \tau_1 \big\}}$, lies obviously in $\Cc(x_0)$. Furthermore,
\beq \label{inter1}
\int_0^{\tau_1} e^{-\rho s} U(c_s) ds &=& \int_0^{\tau_c \wedge \tau_1} e^{-\rho s} U(c'_s) ds + \int_{\tau_c \wedge \tau_1}^{\tau_1} e^{-\rho s}
U(c_s) ds \nonumber \\
&\leq& \int_0^{\tau_1} e^{-\rho s} U(c'_s) ds  \; + \; \frac{U(\rho(x-x_0)_+)}{\rho},
\enq
by the same Jensen's arguments as in \reff{jensen}, and for all $y$ $\geq$ $0$,
\beq \label{inter2}
\hat{v}_{_{I_{\tau_1}^i}}\Big(x - \int_0^{\tau_1} c_t dt + y \frac{S_{\tau_1}}{S_0}\Big) &\leq&
\hat{v}_{_{I_{\tau_1}^i}}\Big(x_0 - \int_0^{\tau_1} c'_t dt + (x - x_0)_+  +  y \frac{S_{\tau_1}}{S_0}\Big) \nonumber \\
&\leq& \hat{v}_{_{I_{\tau_1}^i}}\Big(x_0 - \int_0^{\tau_1} c'_t dt \Big)  + \hat{v}_{_{I_{\tau_1}^i}}\Big((x - x_0)_+ + y \frac{S_{\tau_1}}{S_0}\Big),
\enq
where we have used the fact that $\hat{v}_i$ is nondecreasing, and subadditive (as a concave function with $\hat{v}_i(0) \geq 0$).
By adding  the two inequalities \reff{inter1}-\reff{inter2}, and taking expectation, we obtain from \reff{dynvi1}:
\beqs
v_i(x,y) &\leq& v_i(x_0,0) + \frac{U(\rho(x-x_0)_+)}{\rho} + \E \Big[ e^{-\rho\tau_1}\hat{v}_{_{I_{\tau_1}^i}}\Big((x - x_0)_+ + y \frac{S_{\tau_1}}{S_0}\Big) \Big],
\enqs
and by the same domination arguments as in the first case, this shows that
\beqs
\limsup_{(x,y) \rightarrow (x_0,0)} v_i(x,y) &\leq& v_i(x_0,0),
\enqs
which ends the proof.
\ep

\begin{Remark} \label{rm:proofcontinuity}
{\rm
The above proof of continuity of the value functions at the boundary by means of the dynamic programming principle is somehow different from other similar proofs that one can find e.g. in \cite{DFG11,phatan08,zari02}.
Indeed in such problems the proof of dynamic programming principle is done (or referred to) in two parts: the ``easy'' one ($\le$) which does not require continuity of the value function, and the `difficult'' one ($\ge$) which requires the continuity of the value function up to the boundary. The proof of continuity at the boundary in such cases uses only the ``easy'' inequality.
In our case, due to the specific boundary condition of our problem, the ``easy'' inequality is not enough to prove the continuity at the boundary. We need also the ``hard'' inequality. For this reason we give, in Appendix A, a proof of the dynamic programming principle in our case that, in the ``hard'' inequality part, uses the continuity of $v_i$ in the interior and the continuity of its restriction to the boundary (which are both implied by the concavity and by the growth condition (\ref{growthvi})).
\hfill\qed
}
\end{Remark}

\vspace{2mm}
\begin{Remark} \label{rm:negpower}
{\rm
For simplicity we have restricted our study to the case where $U$ is defined on the positive half-line $\R_+$. With some work, our results can be extended to the case $U(0)=-\infty$, assuming $U(c) \geq - K c^{q}$, for some $K \geq 0$, $q<0$. In that case (assuming $\rho >0$), $v_i(x,y) > - \infty$ whenever $x>0$, $y \geq 0$, while $v_i(0^+,y)=-\infty$ for all $y$.
\hfill\qed
}
\end{Remark}

\vspace{2mm}

We shall also need in Section \ref{sec:pow} the following technical lemma.

\begin{Lemma} \label{lemLowerBoundV_x}
There exists some positive constant  $C >0$ s.t.
\beq \label{eqLowerBoundV_x}
\Dx{v_i} (x^+, y) := \lim_{\delta\downarrow 0} \frac{v_i(x+\delta,y)-v_i(x,y)}{\delta}  &\geq& C \; U'(2x), \;\;\;\; \forall \; x,y \in \R_+, i \in \I_d.
\enq
\end{Lemma}

{\noindent \bf Proof.}  Fix some  $x,y \geq 0$,  and set $x_1 = x + \delta$ for  $\delta>0$. For any  $(\zeta,c)$ $\in$ $\Ac_i(x,y)$ with associated cash/amount in shares  $(X,Y)$, notice that
 $(\tilde{\zeta},\tilde c)$ $:=$ $(\zeta,c + \delta \mathbf{1}_{[0,1 \wedge \tau_1 ]})$ is admissible for $(x_1,y)$. Indeed, the associated  cash amount satisfies
\beqs
\tilde X_t &=& X_t + (x_1 - x) - \int_0^t \delta \mathbf{1}_{[0,1 \wedge \tau_1 ]}(s) ds  \; \geq \; X_t  \; \geq \;  0,
\enqs
while the amount in cash  $\tilde Y_t$ $=$ $Y_t$ $\geq$ $0$ since $\zeta$ is unchanged.  Thus, $(\tilde\zeta,\tilde{c}) \in \Ac_i(x_1,y)$, and we have
\beq
v_i(x_1,y) &\geq& \E\left[ \int_0^\infty e^{-\rho t} U(\tilde{c_t}) dt \right] \nonumber \\
					&=& \E\left[ \int_0^\infty e^{-\rho t} U(c_t) dt \right] + \E\left[ \int_0^{1 \wedge \tau_1} e^{-\rho t} \left(U(c_t + \delta) -U(c_t)\right)  dt \right].  \label{eqMonoX}
\enq
Now, by concavity of $U$: $U(c_t + \delta) - U(c_t) \geq \delta U'(c_t + \delta)$, and
\beq
\int_0^{1 \wedge \tau_1} e^{-\rho t} (U(c_t + \delta) - U(c_t)) dt  &\geq& \int_0^{1 \wedge \tau_1} e^{-\rho t} \delta U'(c_t + \delta) dt \nonumber \\
&\geq& \delta e^{-\rho(1 \wedge \tau_1)} \int_0^{1 \wedge \tau_1} U'(c_t +\delta) dt \nonumber \\
&\geq& \delta e^{-\rho (1 \wedge \tau_1)} U'(2x+\delta) \int_0^{1 \wedge \tau_1} \mathbf{1}_{\left\{c_t < 2x \right\}} dt. \label{Uc1}
\enq
Moreover,
\beqs
2x \int_0^{1 \wedge \tau_1} \mathbf{1}_{\left\{c_t \geq 2x \right\}} dt \leq \int_0^{1 \wedge \tau_1} c_t dt \leq x,
\enqs
since $(\zeta,c)$ is admissible for $(x,y)$, so that
\beq \label{Uc2}
\int_0^{1 \wedge \tau_1} \mathbf{1}_{\left\{c_t < 2x \right\}} dt \geq (1 \wedge \tau_1) - \left(\frac{1}{2} \wedge \tau_1 \right) \geq \frac{1}{2} \mathbf{1}_{\left\{\tau_1 \geq 1 \right\}}.
\enq
By combining \reff{Uc1} and \reff{Uc2}, and taking the expectation, we get
\beqs
\E\left[\int_0^{1 \wedge \tau_1} e^{-\rho t} (U(c_t + \delta) - U(c_t)) dt \right] &\geq& \delta U'(2x+\delta)\E\Big[e^{-\rho(1 \wedge \tau_1)}  \frac{1}{2} \mathbf{1}_{\left\{\tau_1 \geq 1 \right\}} \Big] .
\enqs
By taking the supremum over $(\zeta,c)$ in \reff{eqMonoX}, we thus obtain with the above inequality
\beqs
v_i(x + \delta,y) &\geq& v_i(x,y) + \delta U'(2x+\delta) \E\left[e^{-\rho(1 \wedge \tau_1)}  \frac{1}{2} \mathbf{1}_{\left\{\tau_1 \geq 1 \right\}} \right].
\enqs
Finally, by choosing $C=\E\big[e^{-\rho(1 \wedge \tau_1)}\frac{1}{2} \mathbf{1}_{\left\{\tau_1 \geq 1 \right\}} \big] >0$, and letting $\delta$ go to $0$, we obtain the required inequality \reff{eqLowerBoundV_x}.
\ep

\section{Dynamic programming and viscosity characterization}

\setcounter{equation}{0} \setcounter{Assumption}{0}
\setcounter{Theorem}{0} \setcounter{Proposition}{0}
\setcounter{Corollary}{0} \setcounter{Lemma}{0}
\setcounter{Definition}{0} \setcounter{Remark}{0}

In this section, we provide an analytic characterization of the value functions $v_i$, $i$ $\in$ $\I_d$,  to our control problem
\reff{defVi}, by relying on the dynamic programming principle, which is shown to hold and  formulated  as:

\begin{Proposition} \label{propDPP} (Dynamic programming principle)
For all $(x,y,i)$ $\in$ $\R_+^2\times\I_d$, and any stopping time $\tau$, we have
\beq \label{DPP}
v_i(x,y) &=& \sup_{(\zeta,c)\in\Ac_i(x,y)} \E\Big[ \int_0^\tau e^{-\rho t} U(c_t) dt + e^{-\rho \tau} v_{_{I_\tau}}(X_\tau,Y_\tau)  \Big].
\enq
\end{Proposition}
{\noindent \bf Proof.} See Appendix A.
\ep

\vspace{3mm}

The associated dynamic programming system (also called Hamilton-Jacobi-Bellman or HJB system) for $v_i$, $i$ $\in$ $\I_d$, is written as
\beq
  \rho v_i  -  b_i y \Dy{v_i} \; - \;  \frac{1}{2} \sigma_i^2 y^2 \Dyy{v_i}\; - \; \tilde U\left( \Dx{v_i} \right) & &  \label{eqHJB}  \\
 - \sum_{j\neq i} q_{ij} \Big[  v_j\big(x, y(1-\gamma_{ij})\big) - v_i(x,y)  \Big] & & \nonumber \\
 \; - \; \lambda_i \ \big[ \hat v_i(x+y) - v_i(x,y) \big]  &=& 0, \;\;\;  (x,y) \in (0,\infty)\times\R_+,  \; i \in \I_d, \nonumber
\enq
together with the boundary condition \reff{vix=0} on $\{0\}\times\R_+$ for $v_i$, $i$ $\in$ $\I_d$. Notice that, arguing as one does for the deduction of the HJB system above, the boundary condition \reff{vix=0} may also be written as:
\beq
 \rho v_i(0,.)  -  b_i y \Dy{v_i}(0,.) \; - \;  \frac{1}{2} \sigma_i^2 y^2 \Dyy{v_i}(0,.) & &   \nonumber   \\
 - \sum_{j\neq i} q_{ij} \Big[  v_j\big(0, y(1-\gamma_{ij})\big) - v_i(0,y)  \Big] & & \nonumber \\
 \; - \; \lambda_i   \big[ \hat v_i(y) - v_i(0,y) \big]  &=& 0, \;\;\; y > 0, \; i \in \I_d.  \label{bounvis}
\enq
Notice that in this boundary condition the term  $\tilde U\left( \Dx{v_i} \right)$
has disappeared. This implicitly comes from the fact that, on the boundary $x=0$ the only admissible consumption rate is $c=0$. We will say more on this in studying the case of CRRA utility function in Section 5.1.

\vspace{2mm}

In our context, the notion of  viscosity solution to the non local second-order system $(E)$  is defined as follows.

\begin{Definition}
(i) A d-tuple $w$ $=$ $(w_i)_{i\in\I_d}$ of continuous functions on $\R_+^2$ is a viscosity supersolution (resp. subsolution) to \reff{eqHJB} if
\beqs
 \rho \varphi_i(\bar x,\bar y)  -  b_i \bar y \Dy{\varphi_i}(\bar x,\bar y)  \; - \;  \frac{1}{2} \sigma_i^2 \bar y^2 \Dyy{\varphi_i}(\bar x,\bar y)
 \; - \; \tilde U\left( \Dx{\varphi_i}(\bar x,\bar y)  \right) & &  \nonumber   \\
 - \sum_{j\neq i} q_{ij} \Big[  \varphi_j\big(\bar x, \bar y(1-\gamma_{ij})\big) - \varphi_i(\bar x,\bar y)  \Big] & & \nonumber \\
 \; - \; \lambda_i \ \big[ \hat \varphi_i(\bar x+\bar y) - \varphi_i(\bar x,\bar y) \big]  & \geq \; (\mbox{ resp. } \leq ) & 0,
\enqs
for all d-tuple $\varphi$ $=$ $(\varphi_i)_{i\in\I_d}$ of $C^2$ functions on $\R_+^2$, and any  $(\bar x,\bar y,i)$ $\in$
$(0,\infty)\times\R_+\times\I_d$,  such that $w_i(\bar x,\bar y)$ $=$ $\varphi_i(\bar x,\bar y)$, and $w$ $\geq$ (resp. $\leq$) $\varphi$ on $\R_+^2\times\I_d$.

\vspace{1mm}

\noindent (ii) A d-tuple $w$ $=$ $(w_i)_{i\in\I_d}$ of continuous functions on $\R_+^2$ is a viscosity  solution to \reff{eqHJB} if it is both a viscosity supersolution and subsolution to \reff{eqHJB}.
\end{Definition}

The main result of this section is to provide an analytic characterization of the value functions in terms of viscosity solutions to the dynamic programming system.

\begin{Theorem} \label{thmviscovi}
The value function $v$ $=$  $(v_i)_{i\in\I_d}$ is  the unique viscosity solution to \reff{eqHJB} sa\-tisfying the boundary condition \reff{vix=0}, and the growth condition \reff{growthvi}.
\end{Theorem}
{\bf Proof.}
The proof of viscosity property follows as usual from the dynamic programming principle. The uniqueness and comparison result for viscosity solutions is proved by rather standard arguments, up to some specificities related to the non local terms  and state constraints induced by our hybrid jump-diffusion control problem. We postponed the details in Appendix B.
\ep

\section{The case of CRRA utility } \label{sec:pow}

\setcounter{equation}{0} \setcounter{Assumption}{0}
\setcounter{Theorem}{0} \setcounter{Proposition}{0}
\setcounter{Corollary}{0} \setcounter{Lemma}{0}
\setcounter{Definition}{0} \setcounter{Remark}{0}

In this section, we consider the case where the utility function is of CRRA type in the form:
\beq \label{CRRA}
U(x) &=& \frac{x^p}{p}, \;\;\; x > 0, \;\; \mbox{ for some } \; p \in (0,1).
\enq
We shall exploit the homogeneity property  of  the CRRA utility function,  and  go beyond the viscosity characterization
of the value function in order to prove some regularity results, and provide an explicit characterization of the optimal control through
a verification theorem.  We next give a numerical analysis  for computing the value functions and optimal strategies, and illustrate with some tests for measuring the impact of our illiquidity  features.

\subsection{Regularity results and verification theorem}

For any $(i,x,y)$ $\in$ $\I_d\times\R_+^2$, $(\zeta,c)$ $\in$ $\Ac(x,y)$  with associated  state process $(X,Y)$,  we notice from the dynamics
\reff{dynX}-\reff{dynY} that for any $k$ $\geq$ $0$, the state $(kX,kY)$ is associated to the control $(k\zeta,kc)$.  Thus, for $k>0$,we have $(\zeta,c)$ $\in$
$\Ac_i(x,y)$ iff $(k\zeta,kc)$ $\in$ $\Ac(kx,kc)$, and so from the homogeneity property of the power utility function $U$ in \reff{CRRA}, we have:
\beq \label{vikp}
v_i(kx,ky) &=& k^p v_i(x,y), \;\;\; \forall (i,x,y) \in \I_d\times\R_+^2, \; k \in \R_+.
\enq
Let us now consider the change of variables:
\beqs
(x,y) \in \R_+^2 \setminus\{(0,0)\} & \longrightarrow & \big(r  =   x + y,  z  =   \frac{y}{x+y} \big) \; \in \; (0,\infty) \times [0,1].
\enqs
Then, from \reff{vikp}, we have $v_i(x,y)$ $=$ $v_i(r(1-z),rz)$ $=$ $r^pv_i(1-z,z)$, and we can separate the value function $v_i$ into:
\beq \label{changevar}
v_i(x,y) &=&  U(x+y)  \varphi_i\Big(\frac{y}{x+y}\Big), \;\;\;  \forall (i,x,y) \in \I_d\times ( \R_+^2 \setminus\{(0,0)\} )
\enq
where $\varphi_i(z)$ $=$ $p \; v_i(1-z,z)$ is a continuous function on $[0,1]$.
By substituting this transformation for $v_i$ into the dynamic programming equation \reff{eqHJB} and the boundary condition
\reff{bounvis}, and after some straightforward calculations, we see that  $\varphi$ $=$ $(\varphi_i)_{i\in\I_d}$ should solve the system of (nonlocal)
ordinary differential equations (ODEs):
\beq
(\rho   - p b_i z  + \frac{1}{2}p(1-p) \sigma_i^2 z^2) \varphi_i - (1-p) \Big( \varphi_i-\frac{z}{p} \varphi_i'\Big)^{-\frac{p}{1-p}} &   &  \label{eqHJBPhi}   \\
-   \; z(1-z)(b_i - z(1-p)\sigma_i^2) \varphi_i' - \frac{1}{2} z^2(1-z)^2 \sigma_i^2 \varphi_i''   & & \nonumber \\
- \; \sum_{j\neq i} q_{ij} \Big[  (1-z\gamma_{ij})^p \varphi_j\Big(\frac{z(1-\gamma_{ij})}{1-z\gamma_{ij}}\Big) - \varphi_i(z)  \Big] & & \nonumber \\
 \; - \; \lambda_i \sup_{\pi \in [0,1]} \big[ \varphi_i(\pi) - \varphi_i(z) \big]  &=& 0, \;\;\;   z \in [0,1), \;\; i \in \I_d, \nonumber
\enq
together with the boundary condition for $z$ $=$ $1$:
\beq
(\rho  - p b_i  + \frac{1}{2}p(1-p) \sigma_i^2) \varphi_i(1) & & \nonumber \\
\; - \;  \sum_{j\neq i} q_{ij} \big[ (1- \gamma_{ij})^p \varphi_j(1)-\varphi_i(1)\big]
- \lambda_i \sup_{\pi \in [0,1]} \big[ \varphi_i(\pi) - \varphi_i(1)\big] &=& 0, \;\;\; i \in \I_d.   \label{eqPhiBndry1}
\enq
The following boundary condition for $z$ $=$ $0$, obtained formally by taking $z$ $=$ $0$ in \reff{eqHJBPhi},
\beq
\rho \varphi_i(0)  - (1-p) \big( \varphi_i(0) \big)^{-\frac{p}{1-p}} & & \nonumber \\
\; - \;  \sum_{j\neq i} q_{ij} \big[\varphi_j(0)-\varphi_i(0)\big]
- \lambda_i \sup_{\pi \in [0,1]} \big[ \varphi_i(\pi) - \varphi_i(0)\big] &=& 0,  \;\;\; i \in \I_d, \label{eqPhiBndry0}
\enq
is  proved rigorously in the below Proposition.


\begin{Proposition} \label{propRegPhi}
The $d$-tuple  $\varphi$ $=$ $(\varphi_i)_{i\in\I_d}$ is concave on $[0,1]$, $C^2$ on $(0,1)$.
We further have
\beq
\lim_{z \rightarrow 0} z \varphi'_i(z)  &=& 0, \label{eqlimPhiPrime0} \\
\lim_{z \rightarrow 0} z^2 \varphi''_i(z) &=& 0, \label{eqlimPhiSecond0} \\
\lim_{z \rightarrow 1} (1-z) \varphi'_i(z)  &=& 0, \label{eqlimPhiPrime1b} \\
\lim_{z \rightarrow 1} (1-z)^2 \varphi''_i(z) &=& 0, \label{eqlimPhiSecond1} \\
\lim_{z \rightarrow 1} \varphi'_i(z) &=& - \infty, \label{eqlimPhiPrime1a}
\enq
 and $\varphi$ is the unique bounded classical solution of
\reff{eqHJBPhi} on $(0,1)$, with boundary conditions  \reff{eqPhiBndry1}-\reff{eqPhiBndry0}.
\end{Proposition}
{\bf Proof.} Since  $\varphi_i(z)$ $=$  $p \;  v_i(1-z,z)$, and by concavity of $v_i(.,.)$ in both variables, it is clear that  $\varphi_i$ is concave on $[0,1]$.
From the viscosity property of $v_i$ in Theorem \ref{thmviscovi}, and the change of variables \reff{changevar},  this implies that  $\varphi$ is the unique bounded  viscosity solution to \reff{eqHJBPhi} on $[0,1)$, satisfying the boundary condition  \reff{eqPhiBndry1}.  Now, recalling that $q_{ii}$ $=$ $-\sum_{j\neq i} q_{ij}$, we observe that
the system  \reff{eqHJBPhi} can be written as:
\beq
&& (\rho - q_{ii} + \lambda_i  - p b_i z  + \frac{1}{2}p(1-p) \sigma_i^2 z^2) \varphi_i -  z(1-z)(b_i - z(1-p)\sigma_i^2) \varphi_i' \nonumber \\
 && \; - \;  \frac{1}{2} z^2(1-z)^2 \sigma_i^2 \varphi_i''  - (1-p) \big( \varphi_i-\frac{z}{p} \varphi_i'\big)^{-\frac{p}{1-p}} \nonumber \\
&=& \sum_{j\neq i} q_{ij} \left[  (1-z\gamma_{ij})^p \varphi_j\Big(\frac{z(1-\gamma_{ij})}{1-z\gamma_{ij}}\Big) \right] + \lambda_i \sup_{\pi \in [0,1]} \varphi_i(\pi), \;\;\; z \in (0,1), \;
i \in \I_d.  \label{odevarphi}
\enq
Let us  fix some $i$ $\in$ $\I_d$, and an arbitrary compact $[a,b]$ $\subset$ $(0,1)$. By standard results, see e.g. \cite{craishlio92},  we know that the  second-order ODE:
\beq
&& (\rho - q_{ii} + \lambda_i  - p b_i z  + \frac{1}{2}p(1-p) \sigma_i^2 z^2) w_i -  z(1-z)(b_i - z(1-p)\sigma_i^2) w_i' \nonumber \\
 && - \frac{1}{2} z^2(1-z)^2 \sigma_i^2 w_i''  - (1-p) \big( w_i-\frac{z}{p} w_i'\big)^{-\frac{p}{1-p}} \nonumber \\
&=& \sum_{j\neq i} q_{ij} \left[  (1-z\gamma_{ij})^p \varphi_j\Big(\frac{z(1-\gamma_{ij})}{1-z\gamma_{ij}}\Big) \right] + \lambda_i \sup_{\pi \in [0,1]} \varphi_i(\pi) \label{odew}
\enq
has a unique viscosity solution $w_i$ satisfying $w_i(a)=\varphi_i(a)$, $w_i(b)=\varphi_i(b)$, and that this solution $w_i$ is twice differentiable on $[a,b]$  since
the second term $z(1-z)\sigma_i^2$ is uniformly elliptic on $[a,b]$, see \cite{ladsol68}.
Since $\varphi_i$ is a viscosity solution to \reff{odew} by \reff{odevarphi}, we deduce by uniqueness that $\varphi_i$ $=$ $w_i$ on $[a,b]$. Since $a,b$ are arbitrary, this means that  $\varphi$ is $C^2$ on $(0,1)$.
By concavity of  $\varphi_i$, we have  for all $z$ $\in$ $(0,1)$,
\beqs
\frac{\varphi_i(1) - \varphi_i(z)}{1-z} \leq \varphi'_i(z) \leq \frac{\varphi_i(z) - \varphi_i(0)}{z}.
\enqs
Letting $z$ $\rightarrow$ $0$ and $z$ $\rightarrow$ $1$, and by continuity of $\varphi_i$, we obtain \reff{eqlimPhiPrime0} and \reff{eqlimPhiPrime1b}.

Now letting $z$ go to $0$ in \reff{eqHJBPhi}, we obtain $\lim_{z\rightarrow 0} z^2 \varphi''_i(z) =l$ for some finite $l \leq 0$. If $l <0$, $z^2 \varphi''_i(z) \leq \frac{l}{2}$ whenever $z \leq \eta$, for some $\eta >0$. By writing that
\beqs
z (\varphi'_i(z)- \varphi'_i(\eta)) = z \int_{\eta}^z \varphi''_i(u) du \geq - \frac{l}{2} z \int_{z}^\eta \frac{du}{u^2} = \frac{l}{2} z \left( \frac{1}{\eta} - \frac{1}{z}\right),
\enqs
and sending $z$ $\rightarrow$ $0$, we get $\liminf_{z \rightarrow 0} z \varphi'_i(z) \geq -l/2$, which contradicts \reff{eqlimPhiPrime0}. Thus $l$ $=$ $0$, and the boundary condition \reff{eqPhiBndry0}  follows by letting $z \rightarrow 0$ in \reff{eqHJBPhi}.
In the same way, letting $z$ $\rightarrow$ $1$ in \reff{eqHJBPhi} and comparing with \reff{eqPhiBndry1}, we have
\beqs
\lim_{z \rightarrow 1} \frac{1}{2} (1-z)^2 \varphi''_i(z) &=& \left(\varphi_i(1) - \varphi'_i(1-)\right)^{- \frac{p}{1-p}} \;\; \in [0, \infty].
\enqs
\reff{eqlimPhiPrime1b} implies that this limit is $0$, and we obtain \reff{eqlimPhiSecond1} and \reff{eqlimPhiPrime1a}.
\ep

\begin{Remark} \label{remregulv}
{\rm  From \reff{changevar} and the above Proposition, we deduce that the value functions $v_i$, $i$ $\in$ $\I_d$, are $C^2$ on  $(0,\infty) \times (0,\infty)$, and so
are solutions to the dynamic programming system \reff{eqHJB} on $(0,\infty) \times (0,\infty)$ in classical sense.
\hfill\qed
}
\end{Remark}

\vspace{1mm}

We now provide an explicit construction of the optimal investment/consumption strategies in feedback form in terms of the smooth solution $\varphi$ to \reff{eqHJBPhi}-\reff{eqPhiBndry0}-\reff{eqPhiBndry1}.
We start with the following Lemma.

\begin{Lemma} \label{lemphi}
For any $i$ $\in$ $\I_d$, let us define:
\beqs
c^{*}(i,z) &=& \left\{\begin{array}{ll}
\left( \varphi_i(z) - \frac{z}{p} \varphi_i'(z)\right)^{\frac{-1}{1-p}} & \hbox{ when } 0<z<1\\
\left( \varphi_i(0)\right)^{\frac{-1}{1-p}} & \hbox{ when } z=0\\
0  & \hbox{ when } z=1
\end{array}\right. ,\\
\pi^{*}(i) & \in & \arg \max_{\pi \in [0,1]} \varphi_i(\pi).
\enqs
Then for each $i \in \I_d$, $c^{*}(i,.)$ is continuous on $[0,1]$, $C^1$ on $(0,1)$, and given any initial conditions $(r,z)$ $\in$ $\I_d\times\R_+\times [0,1]$,  there exists a solution $(\hat R_t,\hat Z_t)_{t\geq 0}$ valued in
$\R_+\times [0,1]$ to the SDE:
\beq
\label{sdeFeedbackR}
d\hat R_t &=&   \hat R_{t^-} \hat Z_{t^-} \Big( b_{I_{t^-}}  dt + \sigma_{I_{t^-}}   dW_t -  \gamma_{_{I_{t^-},I_t}} dN^{I_{t^-},I_t}_t \Big) - \hat R_{t-} c^{*}(I_{t-},\hat Z_{t-}) dt, \\
d\hat Z_t &=& \hat Z_{t^-}(1-\hat Z_{t^-}) \Big[  \big(b_{I_{t^-}} - \hat Z_{t^-} \sigma_{I_{t^-}}^2\big) dt  + \sigma_{I_{t^-}} dW_t  -
\frac{\gamma_{_{I_{t^-},I_t}} }{1- \hat Z_{t^-} \gamma_{_{I_{t^-},I_t}} }  dN^{I_{t^-},I_t}_t \Big]  \nonumber \\
& & \;\;\;\;\; + \;  (\pi^*(I_{t-}) - \hat Z_{t-}) dN_t \; + \; \hat Z_t c^{*}(I_{t-},\hat Z_{t-})dt. \label{sdeFeedbackZ}
\enq
Moreover, if $r$ $>$ $0$, then $\hat R_t$ $>$ $0$, a.s. for all $t$ $\geq$ $0$.
\end{Lemma}
{\bf Proof.}  First notice that Lemma \ref{lemLowerBoundV_x}, written in terms of the variables $(r,z)$, is formulated equivalently as
\beqs
\varphi_i(z) - \frac{z}{p} \varphi_i'(z) &\geq& C 2^{p-1} (1-z)^{p-1}, \;\;\; z \in (0,1).
\enqs
This implies that $c^*(i,.)$ is well-defined on $(0,1)$, and $C^1$ since $\varphi$ is $C^2$. The continuity of $c^*(i,.)$ at $0$ and $1$ comes from \reff{eqlimPhiPrime0} and \reff{eqlimPhiPrime1a}.

Let us  show the  existence of a solution $Z$ to the SDE  \reff{sdeFeedbackZ}.  We start by the existence of a solution for $t<\tau_1$ (recall that $(\tau_n)$ is the sequence of jump times of $N$).
In the case where  $z=1$ (resp. $z=0$), then $Z_t \equiv 1$ (resp. $Z_t\equiv 0$) is clearly a solution on $[0,\tau_1)$.  Consider now the case where $z$ $\in$ $(0,1)$.
From the local Lipschitz property of $z \mapsto z c^*(i,z)$, and recalling that $\gamma_{ij}$ $<$ $1$,
we know, adapting e.g. the result of Theorem 38, page 303 of \cite{Protterbook04}, that there exists a solution to
\beq
d\hat Z_t &=& \hat Z_{t^-}(1-\hat Z_{t^-}) \Big[  \big(b_{I_{t^-}} - \hat Z_{t^-} \sigma_{I_{t^-}}^2\big) dt  + \sigma_{I_{t^-}} dW_t  -
\frac{\gamma_{_{I_{t^-},I_t}} }{1- \hat Z_{t^-} \gamma_{_{I_{t^-},I_t}} }  dN^{I_{t^-},I_t}_t \Big]  \nonumber \\
& & \;\;\;\;\;  + \; \hat Z_t c^{*}(I_{t-},\hat Z_{t-})dt, \label{sdeFeedbackZ1}
\enq
which is valued in $[0,1]$ up to time $t<\tau'_1:= \tau_1 \wedge \left( \lim_{\eps \rightarrow 0} \inf \left\{t \geq 0| \hat Z_{t}(1- \hat Z_t) \leq \eps  \right\}\right)$. By noting that  $\hat Z_t \geq Z^0_t$, where
\beqs
Z^0_t &=& \frac{z \frac{S_t}{S_0}}{z \frac{S_t}{S_0} + (1-z)}, \;\;\; t \geq 0,
\enqs
is the solution to \reff{sdeFeedbackZ1} without the consumption term, and since $S$ is locally bounded away from $0$, we have $\lim_{t \rightarrow \tau'_1} Z_t=1$ on $\left\{\tau'_1 < \tau_1 \right\}$.
 By extending  $\hat Z_t \equiv 1$ on $[\tau'_1,\tau_1)$, we obtain actually a  solution on $[0,\tau_1)$. Then at $\tau_1$, by taking
$\hat Z_{\tau_1}$ $=$ $\pi^*(I_{\tau_1-})$,  we obtain a solution to \reff{sdeFeedbackZ} valued in $[0,1]$ on $[0, \tau_1]$.  Next, we obtain similarly a solution to \reff{sdeFeedbackZ} on $[\tau_1, \tau_2]$ starting from $\hat Z_{\tau_1}$. Finally, since  $\tau_n \nearrow  \infty$, a.s.,  by pasting we obtain a solution to \reff{sdeFeedbackZ} for $t \in \R_+$.

Given a solution $\hat Z$ to \reff{sdeFeedbackZ}, the solution $\hat R$ to \reff{sdeFeedbackR} starting from $r$ at time $0$ is determined by  the stochastic exponential:
\beqs
\hat R_t &=&  r \cdot \Ec \left( \int_0^\cdot \hat Z_{s^-} \Big( b_{I_{s^-}}  ds + \sigma_{I_{s^-}}   dW_s -  \gamma_{_{I_{s^-},I_s}} dN^{I_{s^-},I_s}_s \Big) - c^{*}(I_{s-},\hat Z_{s-}) dt \right)_t.
\enqs
Since $- \hat Z_{t-} \gamma_{_{I_{t^-},I_t}} > -1$,  we see that $R_t$ $>$ $0$, $t\geq 0$,  whenever $r$ $>$ $0$, while $R$ $\equiv$ $0$ if $r$ $=$ $0$.
\ep

\vspace{2mm}

\begin{Proposition} \label{propVerif}
Given some initial conditions $(i,x,y)$ $\in$ $\I_d\times(\R_+^2\setminus\{(0,0)\})$, let us consider the pair of processes $(\hat\zeta,\hat c)$ defined by:
\beq
\hat{\zeta_t}   & = & \hat R_{t-}(\pi^{*}(I_{t-}) - \hat Z_{t-}) \\
\hat{c_t} &=& \hat R_{t-}  c^{*}(I_{t-},\hat Z_{t-}),
\enq
where the functions $(c^*,\pi^*)$ are defined in Lemma \ref{lemphi}, and $(\hat R,\hat Z)$ are solutions to \reff{sdeFeedbackR}-\reff{sdeFeedbackZ},
starting from $r$ $=$ $x+y$, $z$ $=$ $y/(x+y)$, with $I$ starting from $i$.  Then,  $(\hat\zeta,\hat c)$ is an optimal  investment/consumption strategy  in $\Ac_i(x,y)$,  with associated state process $(\hat X,\hat Y)$ $=$ $(\hat R(1-\hat Z),\hat R\hat Z)$, for $v_i(x,y)$ $=$  $U(r)\varphi_i(z)$.
\end{Proposition}
{\bf Proof.}  For such choice of $(\hat\zeta,\hat c)$, the dynamics of $(\hat R,\hat Z)$ evolve according to \reff{dynR}-\reff{dynZ} with a feedback
control $(\hat\zeta,\hat c)$, and thus correspond (via It\^o's formula) to a state process $(\hat X,\hat Y)$ $=$ $(\hat R(1-\hat Z),\hat R\hat Z)$ governed by \reff{dynY}-\reff{dynX}, starting from $(x,y)$, and satisfying the nonbankruptcy constraint \reff{eq:nonbankruptcyequivstate}.  Thus,
$(\hat\zeta,\hat c)$ $\in$ $\Ac_i(x,y)$.  Moreover, since $r$ $=$ $x+y$ $>$ $0$,  this implies that $\hat R$ $>$ $0$, and so $(\hat X,\hat Y)$ lies
in $\R_+^2 \setminus\{(0,0)\}$.

As in the proof of the standard verification theorem, we would like to apply It\^o's formula to the function  $e^{-\rho t}v(\hat X_t,\hat Y_t,I_t)$ (denoting by $v(x,y,i)$ $=$ $v_i(x,y)=U(x+y)\varphi_i(y/(x+y))$).
However this is not immediately possible since the process $(\hat X_t,\hat Y_t)$ may reach the boundary of $\R_+^2$ where the derivatives of $v$ do not have classical sense. To overcome this problem, 
we approximate the function $\varphi_i$ (and so $v(x,y,i)$) as follows. 
We define, for every $\varepsilon$ $>$ $0$ a function $\varphi^\varepsilon $ = $(\varphi^\varepsilon)_{i \in \I_d} $ $\in$ $C^2 ([0,1], \R^d)$ as in the proof of Theorem 4.24 in \cite{DFG11}, such that 
\begin{itemize}
	\item $\varphi_i^\eps = \varphi_i$ on $[\eps, 1- \eps]$,
	\item $\varphi_i^\eps \rightarrow \varphi_i$ uniformly on $[0,1]$ as $\eps \rightarrow 0$,
	\item  $z(1-z) (\varphi_i^\eps)' \rightarrow z(1-z) \varphi'_i$ uniformly on $[0,1]$ as $\eps \rightarrow 0$,
	\item  $z^2(1-z)^2 (\varphi_i^\eps)'' \rightarrow z^2(1-z)^2 \varphi''_i$ uniformly on $[0,1]$ as $\eps \rightarrow 0$,
\end{itemize}
Now we can apply Dynkin's formula to the function $v^\varepsilon (x,y,i)=U(x+y)\varphi^\varepsilon_i (y/(x+y)) $ calculated on the process $(\hat X, \hat Y,I)$ between time $0$ and $\tau_n\wedge T$, where $\tau_n$ $=$
$\inf\{t\geq 0: \hat X_t+\hat Y_t  \geq n\}$ :

\beq
v^\eps(x,y,i) &=& \E \Big[  e^{-\rho (\tau_n\wedge T) } v^\eps(\hat X_{\tau_n\wedge T},\hat Y_{\tau_n\wedge T},I_{\tau_n\wedge T}) \nonumber \\
& & \;\;  + \int_0^{\tau_n\wedge T} e^{-\rho t} \Big(  \rho v^\eps +   \hat c_t  \Dx{v^\eps}  -   b_{I_{t^-}} \hat Y_{t^-} \Dy{v^\eps}
-  \frac{1}{2}\sigma_{I_{t^-}}^2 \hat Y_{t^-}^2 \Dyy{v^\eps} \nonumber  \\ 
& & \;\;\;\;\;\;\;\;\;   - \sum_{j\neq I_{t^-} } q_{_{I_{t^-}j}} [ v^\eps(\hat X_{t^-},\hat Y_{t^-}(1-\gamma_{_{I_{t^-}j}}),j) - v^\eps(\hat X_{t^-},\hat Y_{t^-},I_{t^-})]
\nonumber \\
& &  \;\;\;\;\;\;\;\;\;   - \lambda_{_{I_{t^-}}} \big[ v^\eps(\hat X_{t^-} - \hat\zeta_t,\hat Y_{t^-}+\hat\zeta_t,I_{t^-})
- v^\eps(\hat X_{t^-},\hat Y_{t^-},I_{t^-}) \big] \Big) dt \Big]  \label{Dynkin-veps}
\enq
We denote by  $\hat\zeta(i,r,z)$ $=$ $r(\pi^*(i) -z)$, $\hat c(i,r,z)$ $=$ $r c^*(i,z)$, and define $g^\eps$ on $(\R_+^2 \setminus\{(0,0)\})\times \I_d$ by
\beqs
 \rho v^\eps_i  -  b_i y \Dy{v^\eps_i}  -   \frac{1}{2} \sigma_i^2 y^2 \Dyy{v^\eps_i}  +   \hat c(i,x+y,\frac{y}{x+y}) \Dx{v^\eps_i}    -  U\big(\hat c(i,x+y,\frac{y}{x+y})\big)
& &   \label{eqviopt}   \\
 - \sum_{j\neq i} q_{ij} \Big[  v^\eps_j\big(x, y(1-\gamma_{ij})\big) - v^\eps_i(x,y)  \Big] & & \nonumber \\
 \; - \; \lambda_i  \Big[ v^\eps_i\Big( x - \hat\zeta\big(i,x+y,\frac{y}{x+y}\big), y +  \hat\zeta\big(i,x+y,\frac{y}{x+y}\big) \Big)- v^\eps_i(x,y) \Big]   &=:& g_i^\eps(x,y), \nonumber
\enqs
so that from \reff{Dynkin-veps}: 
\beq
v^\eps(i,x,y) & = &  \E \Big[  e^{-\rho (\tau_n\wedge T) } v^\eps(\hat X_{\tau_n\wedge T},\hat Y_{\tau_n\wedge T},I_{\tau_n\wedge T}) \nonumber \\
&& \;\;\;\;\; + \;  \int_0^{\tau_n\wedge T} e^{-\rho t} (U(\hat c_t) + g^\eps(\hat X_t,\hat Y_t, I_t)) dt \Big].  \label{rhs}
\enq
Notice that the properties of $\varphi^\eps$ imply :
\begin{itemize}
	\item $v^\eps_i=v_i$ on $\left\{ \eps \leq \frac{y}{x+y} \leq 1-\eps \right\}$, 
	\item $v_i^\eps \rightarrow v_i$ uniformly on bounded subsets of $\R_+^2$,
	\item $\hat c(i,x+y,\frac{y}{x+y}) \Dx{v^\eps_i} \rightarrow \left\{\begin{array}{ll}
c(i,x+y,\frac{y}{x+y}) \Dx{v_i}, &  x>0\\
0, &  x=0
\end{array}\right. $ uniformly on bounded subsets of $\R_+^2$,
\item $y \Dy{v^\eps_i} \rightarrow \left\{\begin{array}{ll}
y \Dy{v_i}, &  y>0\\
0, &  y=0
\end{array}\right. $ uniformly on bounded subsets of $\R_+^2$,
\item $y^2 \Dyy{v^\eps_i} \rightarrow \left\{\begin{array}{ll}
y^2 \Dyy{v_i}, &  y>0\\
0, &  y=0\end{array}\right. $ uniformly on bounded subsets of $\R_+^2$.
\end{itemize}
The details  can be found in \cite{phdgas11}.  
Since $v$ is a classical solution of \reff{eqHJB} on $(0,\infty) \times (0,\infty)$, this implies that $g^\eps$ converges to $0$ uniformly on bounded subsets of $\R_+^2$ when $\eps$ goes to $0$.
We then obtain  by letting $\eps \to 0$ in \reff{rhs}: 
\beqs
v(x,y,i) &=&  \E \Big[  e^{-\rho (\tau_n\wedge T) } v(\hat X_{\tau_n\wedge T},\hat Y_{\tau_n\wedge T},I_{\tau_n\wedge T})  + \int_0^{\tau_n\wedge T} e^{-\rho t} U(\hat c_t) dt \Big],  
\enqs
From the growth condition \reff{growthvi} we get
\beqs
\E \Big[  e^{-\rho (\tau_n\wedge T) } v(\hat X_{\tau_n\wedge T},\hat Y_{\tau_n\wedge T},I_{\tau_n\wedge T})\Big]
&\le& C \E \Big[ e^{-\rho (\tau_n\wedge T) } R^p_{\tau_n\wedge T} \Big].
\enqs
So, using Lemma \ref{lemGrowthR}, sending $n$ to infinity, and then $T$ to infinity,  we get
\beqs
\lim_{T\rightarrow\infty} \lim_{n\rightarrow\infty}  \E \Big[  e^{-\rho (\tau_n \wedge T) } v(\hat X_{\tau_n\wedge T},\hat Y_{\tau_n\wedge T},I_{\tau_n\wedge T})\Big] &=& 0.
\enqs
Applying monotone convergence theorem to the second term in the r.h.s. of \reff{rhs}, we then obtain
\beqs
v_i(x,y) &=&   \E \Big[  \int_0^{\infty} e^{-\rho t} U(\hat c_t) dt \Big],
\enqs
which proves the optimality of $(\hat\zeta,\hat c)$.
\ep

\subsection{Numerical analysis}

We focus on the numerical resolution of the system of ODEs \reff{eqHJBPhi}-\reff{eqPhiBndry0}-\reff{eqPhiBndry1} satisfied by
$(\varphi_i)_{i\in\I_d}$, and rewritten for all $i$ $\in$ $\I_d$ as:
\beqs
&& (\rho - q_{ii} + \lambda_i  - p b_i z  + \frac{1}{2}p(1-p) \sigma_i^2 z^2) \varphi_i -  z(1-z)(b_i - z(1-p)\sigma_i^2) \varphi_i' \nonumber \\
 && \; - \;  \frac{1}{2} z^2(1-z)^2 \sigma_i^2 \varphi_i''  - (1-p) \big( \varphi_i-\frac{z}{p} \varphi_i'\big)^{-\frac{p}{1-p}} \nonumber \\
&=& \sum_{j\neq i} q_{ij} \left[  (1-z\gamma_{ij})^p \varphi_j\Big(\frac{z(1-\gamma_{ij})}{1-z\gamma_{ij}}\Big) \right] + \lambda_i \sup_{\pi \in [0,1]} \varphi_i(\pi), \;\;\; z \in (0,1),
\enqs
\beqs
 (\rho - q_{ii} + \lambda_i)\varphi_i(0)  - (1-p)  \varphi_i(0)^{-\frac{p}{1-p}}
&=& \sum_{j\neq i} q_{ij}  \varphi_j(0) \;   +  \; \lambda_i \sup_{\pi \in [0,1]} \varphi_i(\pi),   \\
 (\rho  -q_{ii} + \lambda_i - p b_i  + \frac{1}{2}p(1-p) \sigma_i^2) \varphi_i(1)
&=&  \sum_{j\neq i} q_{ij} (1- \gamma_{ij})^p \varphi_j(1) +  \lambda_i \sup_{\pi \in [0,1]}  \varphi_i(\pi).
\enqs

We shall adopt an iterative method to solve this system of integro-ODEs :
 starting with $\varphi^0$ $=$ $(\varphi_i^0)_{i\in\I_d}$ $=$ $0$,  we solve $\varphi^{n+1}$ $=$ $(\varphi_i^{n+1})_{i\in\I_d}$
 as the (classical) solution to the local ODEs where the non local terms are calculated from  $(\varphi^n_i)$ :
\beqs \label{OdePhi_n}
(\rho - q_{ii} + \lambda_i  - p b_i z  + \frac{1}{2}p(1-p) \sigma_i^2 z^2) \varphi_i^{n+1} -  z(1-z)(b_i - z(1-p)\sigma_i^2) (\varphi_i^{n+1})' \nonumber \\ - \frac{1}{2} z^2(1-z)^2 \sigma_i^2 (\varphi_i^{n+1})''
- (1-p) \big( \varphi_i^{n+1}-\frac{z}{p} (\varphi_i^{n+1})'\big)^{-\frac{p}{1-p}} \\
= \; \sum_{j\neq i} q_{ij} \left[  (1-z\gamma_{ij})^p \varphi_j^{n}\Big(\frac{z(1-\gamma_{ij})}{1-z\gamma_{ij}}\Big) \right]
 \; + \; \lambda_i \sup_{\pi \in [0,1]} \varphi_i^{n}(\pi), \nonumber
\enqs
with boundary conditions
\beqs
(\rho - q_{ii} + \lambda_i) \varphi_i^{n+1}(0) - (1-p)\varphi_i^{n+1}(0)^{-\frac{p}{1-p}} &=& \sum_{j\neq i} q_{ij} \varphi_j^{n}(0)
+ \lambda_i \sup_{\pi \in [0,1]} \varphi_i^{n}(\pi) , \\
(\rho - q_{ii} + \lambda_i  - p b_i  + \frac{1}{2}p(1-p) \sigma_i^2) \varphi_i^{n+1}(1) &=&  \sum_{j\neq i} q_{ij} (1- \gamma_{ij})^p \varphi_j^{n}(1) + \lambda_i \sup_{\pi \in [0,1]} \varphi_i^{n}(\pi).
\enqs

Let us denote by:
\beqs
v_i^n(x,y) &=&  \left\{ \begin{array}{cl}
			 U(x+y)  \varphi_i^n\Big(\frac{y}{x+y}\Big), & \;\;\;  \mbox{ for }  (i,x,y) \in \I_d\times ( \R_+^2 \setminus\{(0,0)\} ) \\
			 0, & \;\;\; \mbox{ for }  i \in \I_d, \; (x,y) = (0,0).
			 \end{array}
			 \right.
\enqs
A straightforward calculation shows that $v^n$ $=$ $(v_i^n)_{i\in\I_d}$ are solutions to the iterative local PDEs:
\beq
& &  (\rho -q_{ii} + \lambda_i) v_i^{n+1}  -  b_i y \Dy{v_i^{n+1}} \; - \;  \frac{1}{2} \sigma_i^2 y^2 \Dyy{v_i^{n+1}}
\; - \; \tilde U\big( \Dx{v_i^{n+1}} \big)   \nonumber   \\
&=&   \sum_{j\neq i} q_{ij}  v_j^n\big(x, y(1-\gamma_{ij})\big)
 \; +  \; \lambda_i \hat v_i^n(x+y), \;\;\;  (x,y) \in (0,\infty)\times\R_+,  \; i \in \I_d, \label{PDEvn}
\enq
together with the boundary condition \reff{vix=0} on $\{0\}\times (0,\infty)$ for $v_i$, $i$ $\in$ $\I_d$:
\beq
& &  (\rho- q_{ii} + \lambda_i)  v_i^{n+1}(0,.)  -  b_i y \Dy{v_i^{n+1}}(0,.) \; - \;  \frac{1}{2} \sigma_i^2 y^2 \Dyy{v_i^{n+1}}(0,.) \nonumber \\
&=&   \sum_{j\neq i} q_{ij}  v_j^n\big(0, y(1-\gamma_{ij})\big)  \; + \;
\lambda_i   \hat v_i^n(y), \;\;\; y > 0, \; i \in \I_d. \label{Bndryvn}
\enq

We then have the stochastic control representation for $v^n$ (and so for $\varphi^n$).

\begin{Proposition}
For all $n$ $\geq$ $0$, we have
\beq \label{repvn}
v_i^n(x,y) &=& \sup_{(\zeta,c)\in\Ac_i(x,y)} \E \Big[ \int_0^{\theta_n} e^{-\rho t} U(c_t) dt \Big], \;\;\; (i,x,y) \in \I_d\times\R_+^2,
\enq
where the sequence of random times $(\theta_n)_{n\geq 0}$ are defined by induction  from $\theta_0$ $=$ $0$, and:
\beqs
\theta_{n+1} &=& \inf \left\{t > \theta_n :  \Delta N_t \neq 0 \mbox{ or } \Delta N_t^{I_{t-},I_t} \neq 0 \right\},
\enqs
i.e. $\theta_n$ is the $n$-th time where we have either a change of regime or a trading time.
\end{Proposition}
{\bf Proof.}
Denoting by $w_i^n(x,y)$ the r.h.s. of \reff{repvn}, we need to show that  $w_i^n=v_i^n$. 
First (with a similar proof to Proposition \ref{propDPP}) we have the following Dynamic Programming Principle for the $w^n$ :  for each finite stopping time $\tau$,
\beq \label{dppvn-tau}
w^{n+1}_i(x,y) &=& \sup_{(\zeta,c)\in\Ac_i(x,y)} \E\left[ \int_0^{\tau \wedge \theta_1} e^{-\rho t} U(c_t) dt
+ \mathbf{1}_{\{\tau \geq \theta_1\}} e^{-\rho \theta_1} w^n_{_{I_{\theta_1}}}\left(X_{\theta_1},Y_{\theta_1}\right) \right. \nonumber \\
&& \hspace{25mm} \left. + \mathbf{1}_{\{\tau<\theta_1\}} e^{-\rho \tau} w^{n+1}_{_{I_{\tau}}}\left(X_{\tau},Y_{\tau}\right)   \right]
\enq
The only difference with the statement of  Proposition \ref{propDPP} is the fact that when $\tau\geq\theta_1$, we substitute $w^{n+1}$ with $w^n$ since there are only $n$ stopping times remaining before consumption is stopped 
due to  the finiteness of the horizon in the definition of $w^n$.   

By using \reff{dppvn-tau}, we can show as in Theorem \ref{thmviscovi} that $w^n$ is the unique viscosity solution to \reff{PDEvn}, satisfying boundary condition \reff{Bndryvn} and growth condition \reff{growthvi} (it is actually easier since there are only local terms in this case). Since we already know that $v^n$ is such a solution, it follows that $w^n=v^n$.
\ep

\vspace{2mm}

As a consequence, we obtain the following convergence result for the sequence $(v^n)_n$.

\begin{Proposition}
The sequence $(v^n)_n$  converges increasingly to $v$, and there exists some positive constants $C$ and $\delta$ $<$ $1$ s.t.
\beq \label{vnconv}
0 \; \leq \; v_i - v_i^n & \leq & C  \delta^n (x+y)^p, \;\;\; \forall (i,x,y) \in \I_d\times\R_+^2.
\enq
\end{Proposition}
{\bf Proof.}  First let us show that
\beq \label{IneqDelta}
\delta & := & \sup_{\tiny{\begin{array}{cc}(c,\zeta) \in \Ac_i(x,y) \\ \{(x,y) \in\R_+^2: x+y =1\}\end{array}} }
\E\Big[e^{-\rho \theta_1} R_{\theta_1}^p\Big] \; < \; 1.
\enq
By writing that $e^{-\rho t}R_t^p$ $=$ $D_tL_t$, where $(L_t)_t$ $=$ $(e^{-k(p)t}R_t^p)_t$ is a nonnegative supermartingale by Lemma \ref{lemGrowthR}, and  $(D_t)_t$ $=$ $(e^{-(\rho-k(p))t})_t$ is a decreasing process, we see that $(e^{-\rho t}R_t^p)_t$ is also a  nonnegative supermartingale for all $(\zeta,c)$ $\in$ $\Ac_i(x,y)$, and so:

\beqs
\E\left[e^{-\rho \theta_1}  R^p_{\theta_1} \right] &\leq& \E\left[e^{-\rho (\theta_1 \wedge 1)} R^p_{\theta_1 \wedge 1} \right] \\
&=& \E\left[e^{-(\rho - k(p)) (\theta_1 \wedge 1)} e^{-k(p)(\theta_1 \wedge 1)} R^p_{\theta_1 \wedge 1} \right].
\enqs
Now, since $e^{-(\rho - k(p)) (\theta_1 \wedge 1)}$ $<$ $1$ a.s., $\E\left[e^{-k(p)(\theta_1 \wedge 1)}  R^p_{\theta_1 \wedge 1}\right]$ $\leq$ $1$, for all $(\zeta,c)$ $\in$ $\Ac_i(x,y)$ with $x+y$ $=$ $1$
(recall the supermartingale property of $(e^{-k(p)t}R_t^p)_t$), and by using also the uniform integrability of the family $\left(e^{-k(p)(\theta_1 \wedge 1)} R^p_{\theta_1 \wedge 1}\right)_{c,\zeta}$  from
Lemma  \ref{lemGrowthR}, we obtain the relation \reff{IneqDelta}.

The nondecreasing property of the sequence $(v^n_i)_n$ follows immediately from the representation \reff{repvn}, and we have:
$v_i^n$ $\leq$ $v_i^{n+1}$ $\leq$ $v$ for all $n \geq 0$.  Moreover, the dynamic programming principle \reff{dppvn-tau} applied to $\tau=\theta_1$ gives
\beq \label{DPPvn}
v^{n+1}_i(x,y) &=& \sup_{(\zeta,c)\in\Ac_i(x,y)} \E\left[ \int_0^{\theta_1} e^{-\rho t} U(c_t) dt
+ e^{-\rho \theta_1} v^n_{_{I_{\theta_1}}}\left(X_{\theta_1},Y_{\theta_1}\right)   \right]
\enq
Let us  show \reff{vnconv} by induction on $n$. The case $n=0$ is simply the growth condition \reff{growthvi} since $v^0$ $=$ $0$.
Assume now that \reff{vnconv} holds true at step $n$. From the dynamic programming principle \reff{DPP} and \reff{DPPvn} for $v$ and
$v^{n+1}$, we then have:
\beqs
v^{n+1}_i(x,y)  &\geq& v_i(x,y) - \sup_{(\zeta,c)\in\Ac_i(x,y)} \E\Big[e^{-\rho \theta_1} (v_{I_{\theta_1}}
- v^n_{I_{\theta_1}})\big(X_{\theta_1},Y_{\theta_1}\big)   \Big] \\
&\geq& v_i(x,y) - \sup_{(\zeta,c)\in\Ac_i(x,y)} \E\Big[e^{-\rho \theta_1} C \delta^n R^p_{\theta_1}   \Big] \\
& = & v_i(x,y) - C \delta^{n+1} (x+y)^p,
\enqs
by definition of $\delta$.  This proves the required inequality at step $n+1$, and ends the proof.
\ep

\vspace{2mm}

In the next section, we solve the local ODEs for $\varphi^n$ with Newton's method by a finite-difference scheme (see section 3.2 in \cite{kel92}).

\subsection{Numerical illustrations}

\subsubsection{Single-regime case}

In this paragraph, we consider the case where there is only one regime ($d = 1$).  In this case, our model is similar to the one studied in
\cite{phatan08}, with the key difference that in their model, the investor only observes the stock price at the trading times, so that the consumption process is piecewise-deterministic. We want to compare our results with \cite{phatan08}, and  take the same values for our parameters :  $p=0.5$, $\rho = 0.2$, $b=0.4$, $\sigma = 1$.

Let us recall from \cite{phatan08} the reason behind this choice of parameters (which are not very realistic for a typical financial asset) : to allow meaningful comparison to the Merton (liquid) problem, the optimal Merton investment proportion should be in $[0,1]$, while the liquid value function $v_M$ should be significantly higher than the value function $v_0$ corresponding to the consumption problem without trading. These two constraints correspond to a high risk-return market. In the next subsection (multi-regime case), the choice of parameters will also follow from this reasoning.

Defining the {\it cost of liquidity}  $P(x)$ as the extra amount needed to have the same utility as in the Merton case : $v(x+P(x))$ $=$ $v_M(x)$, we compare the results in our model and in the discrete observation model in \cite{phatan08}. The results in Table 1 indicate that the impact of the lack of continuous observation is quite large, and more important than the constraint of only being able to trade at discrete times.

\begin{table}[h]
\centering
\begin{tabular}{|c|c|c|}
\hline
 $\lambda$ & Discrete observation & Continuous observation \\
\hline
1 & 0.275 & 0.153 \\
5 & 0.121 & 0.016 \\
40 & 0.054 & 0.001 \\
\hline
\end{tabular}
\caption{Cost of liquidity $P(1)$ as a function of $\lambda$.}
\end{table}

In Figure 1 we have plotted the graph of $\varphi(z)$ (actually $\varphi^n(z)$ for $n$ large) and of the optimal consumption rate $c^*(z)$ for different values of $\lambda$. Notice how the value function, the optimal proportion and the optimal consumption rate converge to the Merton values when $\lambda$ increases.

We observe that the optimal investment proportion is increasing with $\lambda$. When $z$ is close to $1$ i.e. the cash proportion in the portfolio is small, the investor faces the risk of  ``having nothing more to consume" and the further away the next trading date is the smaller the consumption rate should be, i.e. $c^*$ is increasing in $\lambda$. When $z$ is far from $1$ it is the opposite : when $\lambda$ is smaller the investor will not be able to invest optimally to maximize future income and should consume more quickly.

\begin{figure}[h]
\include{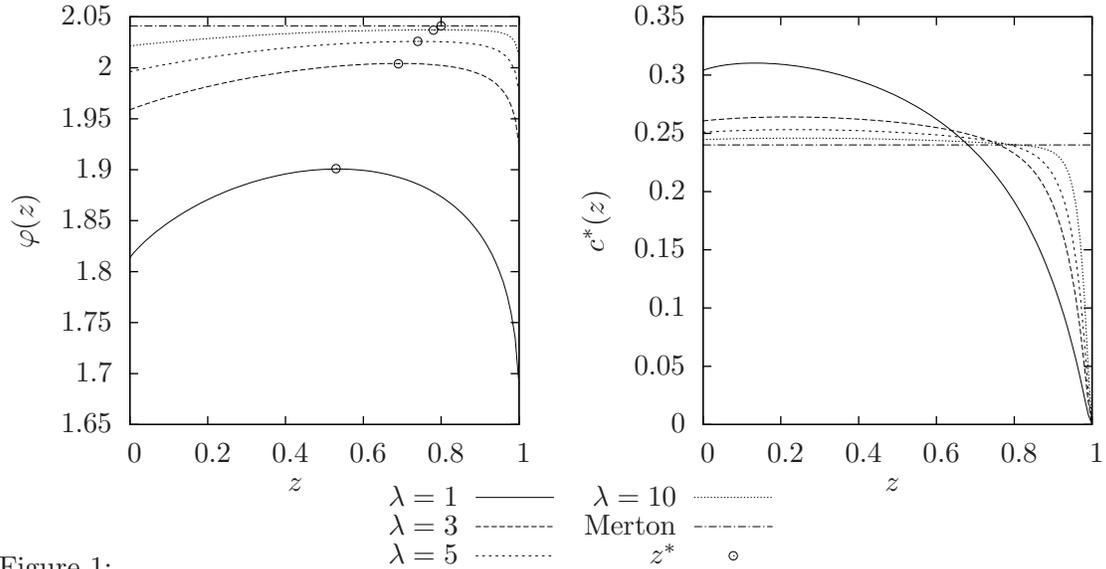}
\caption{\small{ \\ Value function $\varphi(z)$ (left) and optimal consumption rate $c^*(z)$(right) for different values of $\lambda$}} 
\end{figure}

\subsubsection{Two regimes}

\begin{figure}[h!]
\include{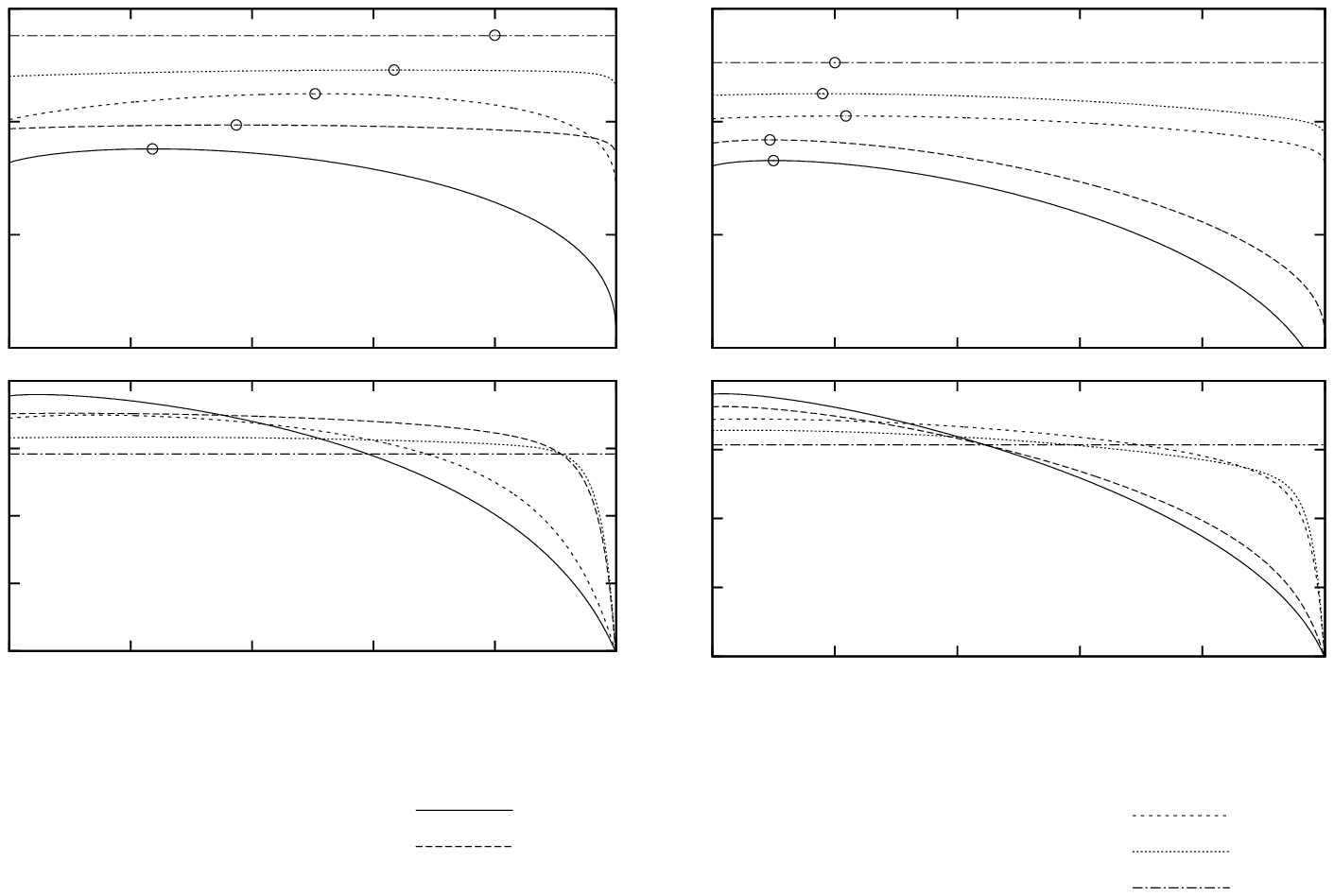}
\caption{$\varphi_i$ and $c^*_i$ for different values of $(\lambda_1,\lambda_2)$}%
\label{fig_2reg}%
\end{figure}

In this paragraph, we consider the case of $d$ $=$ $2$ regimes. We  assume that the asset price is continuous, i.e. $\gamma_{12}$ $=$
$\gamma_{21}$ $=$ $0$.
In this case, the value functions and optimal strategies for the continuous trading (Merton) problem are explicit, see \cite{sotcad09}:
$v_{i,M}(r)$ $=$ $\frac{r^p}{p} \varphi_{i,M}$ where $(\varphi_{i,M})_{i=1,2}$ is the only positive solution to the equations:
\beqs
\Big(\rho - q_{ii} - \frac{b_i^2 p}{2 \sigma_i^2 (1-p)}\Big) \varphi_{i,M} - (1-p) \varphi_{i,M}^{- \frac{p}{1-p}} &=&  q_{ij} \varphi_{j,M}, \;\;\; i,j \in \{1,2\},
\; i \neq j.
\enqs
The optimal proportion invested in the asset $\pi^*_{i,M}= \frac{b_i}{(1-p) \sigma_i^2}$ is the same as in the single-regime case, and the optimal consumption rate is $c^*_{i,M} = (\varphi_{i,M})^{-\frac{1}{p}}$.
We take for values of the parameters
\beqs
p &=& 0.5, \\
q_{12} = q_{21} &=& 1, \\
b_1 = b_2 &=& 0.4, \\
\sigma_1 = 1, &&  \sigma_2 = 2,
\enqs
i.e. the difference between the two market regimes is the volatility of the asset. In Figure \ref{fig_2reg}, we plot the value function and optimal consumption for each of the two regimes in this market, for various values of the liquidity parameters $(\lambda_1,\lambda_2)$. As in the single-regime case, when the liquidity increases,
$\varphi$ and $c^*$ converge to the Merton value.

Note that while in the single regime-case the optimal investment proportion is usually increasing with the liquidity parameter $\lambda$, in the presence of several regimes there does not appear to be a simple similar effect, as can be seen for instance in the upper-right panel of Figure \ref{fig_2reg}.

To quantify the impact of regime-switching on the investor, it is also interesting to compare the cost of liquidity with the single-regime case, see Tables \ref{tbl_dr} and \ref{tbl_sr}. We observe that, for equivalent trading intensity, the cost of liquidity is higher in the regime-switching case. This is economically intuitive : in each regime the optimal investment proportion is different, so that the investor needs to rebalance his portfolio more often (at every change of regime).

\section{Conclusion} 

In this paper we proposed a simple model of an illiquid market with regime-switching, in which the investor may only trade at discrete times corresponding to the arrival times of a Cox process. In this context, we studied an investment/consumption problem over an infinite horizon. In the general case, we proved that the value function for this problem is characterized as the unique viscosity solution to the HJB equation (which is a system of integro-PDEs). In the case of power utility, we proved the regularity of our value function and we were able to characterize the optimal policies. Finally we have presented some numerical results in this special case.

With some straightforward modifications, our viscosity results could be extended to more general regime-switching diffusions (assuming e.g. Lipschitz coefficients). However, the dimension reduction in the case of power utility which allowed us to prove regularity, and made the numerical resolution easier, is specific to our (regime-switching) Black Scholes dynamics.
\newpage
\vspace{5mm}
\hspace{5mm}
\begin{minipage}{0.4\linewidth}
\centering
\begin{tabular}{|c|c|c|}
\hline
 $(\lambda_1,\lambda_2)$ & $P_1(1)$ & $P_2(1)$ \\
\hline
(1,1) & 0.257 & 0.224 \\
(5,5) & 0.112 & 0.103 \\
(10,10) & 0.069 & 0.064 \\
\hline
\end{tabular}
\captionof{table}{\small{Cost of liquidity $P_i(1)$ as a function of $(\lambda_1,\lambda_2)$.}}\label{tbl_dr}
\end{minipage}
\hspace{10mm}
\begin{minipage}{0.4\linewidth}
\centering
\begin{tabular}{|c|c|c|}
\hline
 $\lambda$ & $P_1(1)$& $P_2(1)$ \\
\hline
1 & 0.153 & 0.087 \\
5	& 0.015	&	0.042	\\
10 & 0.004 & 0.024 \\
\hline
\end{tabular}
\captionof{table}{\small{Cost of liquidity $P_i(1)$ for the single-regime case.}}\label{tbl_sr}
\end{minipage}

%
%
%
%
%
%


\newpage

\newpage

\appendix

\section*{Appendix A: Dynamic Programming Principle}

\renewcommand{\theLemma}{A.\arabic{Lemma}}
\renewcommand{\theDefinition}{A.\arabic{Definition}}
\renewcommand{\theProposition}{A.\arabic{Proposition}}
\renewcommand{\theequation}{A.\arabic{equation}}
\renewcommand{\theRemark}{A.\arabic{Remark}}
\renewcommand{\thesection}{A.\arabic{section}}

\setcounter{section}{0}
\setcounter{equation}{0}
\setcounter{Lemma}{0}
\setcounter{Definition}{0}
\setcounter{Proposition}{0}
\setcounter{Remark}{0}


We introduce the weak formulation of the control problem.

\begin{Definition} \label{defWeakControl}
Given $(i,x,y)$ $\in \I_d \times \R_+ \times \R_+$, a control $\Uc$ is a 9-tuple\\ $(\Omega, \Fc, \P, \F=(\Fc_t)_{t\ge 0}, W, I, N, c, \zeta)$, where :
\begin{enumerate}
	\item $(\Omega, \Fc, \P, \F)$ is a filtered probability space satisfying the usual conditions.
	\item $I$ is a Markov chain with space state $\I_d$ and generator $Q$, $I_0=i$ a.s., $N$ is a Cox process with intensity $(\lambda_{I_t})$, and $W$ is an $\F$-Brownian motion independent of $(I,N)$.
	\item $\Fc_t = \sigma(W_s, I_s, N_s; s\leq t) \vee \Nc$, where $\Nc$ is the collection of all $\P$-null sets of $\Fc$.
	\item $(c_t)$ is $\F$-progressively measurable, $(\zeta_t)$ is $\F$-predictable.
\end{enumerate}

We say that $\Uc$ is \emph{admissible}, (writing $\Uc \in \Ac^w_i(x,y)$), if the solution $(X,Y)$ to \reff{dynX}-\reff{dynY} with $X_0=x, Y_0=y$, satisfies $X_t \geq 0$, $Y_t \geq 0$ a.s.
\end{Definition}

Given $\Uc \in \Ac_i^w(x,y)$, define $J(\Uc) = \E\left[ \int_0^\infty e^{-\rho s} U(c_s) ds \right]$, and the value function
\beqs
v_i(x,y) = \sup_{\Uc \in \Ac_i^w(x,y)} J(\Uc).
\enqs

\begin{Proposition} \label{propDPPA}
For every finite stopping time $\tau$ and initial conditions $i,x,y$,
\beq \label{DPP2}
v_i(x,y) &=& \sup_{(\zeta,c)\in\Ac_i^w(x,y)} \E\left[ \int_0^\tau e^{-\rho t} U(c_t) dt + e^{-\rho \tau} v_{I_\tau}(X_\tau,Y_\tau)  \right].
\enq
\end{Proposition}

Before proving this proposition we state some technical lemmas.

\begin{Lemma} \label{lemFtoF0}
Given $(\Omega, \Fc, \P, \F=(\Fc_t), W, I, N)$ satisfying the conditions of Definition \ref{defWeakControl}, define $\F^0 = (\Fc^0_t)_{t\geq 0}$, where $\Fc^0_t= \sigma(W_s,I_s,N_s; s\leq t)$. Then if $(c_t)$ is $\F$-progressively measurable (resp. predictable), there exists $c_1$ $\F^0$-progressively measurable (resp. predictable) such that $c=c_1$ $d\P \otimes dt$ a.e..
\end{Lemma}
{\bf \noindent Proof.} We only give a sketch as the arguments is standard.
We first use Lemma 3.2.4 page 133 in \cite{KS88} to find, for each $n \in \N$,
an approximating ${\Fc}_t$-simple process $c^n$ converging to $c$ in the $L^2 (dt \otimes d\mathbb{P}) $ norm.
Then, using Lemma 1.25 page 13 in \cite{Kallenberg02},  we can change every $c^n$ on a null-set and find a sequence of ${\Fc}_s^{t,0}$-simple process $c_1^n(t)$ that again converges to
$c$ in the $L^2 (dt \otimes d\P) $ norm.
We now extract a subsequence (denoted again by $c_1^n$) such that $c_1^n \to c$ a.e. and we define $c_1 := \lim\inf_{n\to + \infty} c_1^n$. This is $\mathcal{F}_{s}^{t,0}$-progressively measurable and $c=c_{1}$,
 $dt \otimes d\mathbb{P}$ a.e. on $[0,+\infty)\times \Omega$. This concludes the proof.
\ep


\begin{Remark} \label{remF0}
{\rm With the notations of the previous lemma, it is easy to check that $(X^{c',\zeta'},Y^{c',\zeta'})$ $\sim$ $(X^{c,\zeta},Y^{c,\zeta})$ in law. Hence without loss
of generality we can assume that $c$ is $\F^0$-progressively measurable and $\zeta$ is $\F^0$-predictable.
\hfill\qed
}
\end{Remark}

Define $\Wc$ as the space of continuous functions on $\R_+$, $\Ic$ the space of cadlag $\I_d$-valued functions, $\Nc$ the space of nondecreasing cadlag $\N$-valued functions.
On $\Wc \times \Ic \times \Nc$, define the filtration $(\Bc^0_t)_{t \geq 0}$, where $\Bc^0_t$ is the smallest $\sigma$-algebra making the coordinate mappings for $s \leq t$ measurable, and define $\Bc^0_{t+} = \bigcap_{s > t} \Bc^0_s$.

\begin{Lemma} \label{lemCanonical}
If $c$ is $\F^0$-progressively measurable (resp. $\F^0$-predictable), there exists a $\Bc^0_{t+}$-progressively measurable (resp. $\Bc^0_t$-predictable) process $f_c : \R_+ \times \Wc \times \Ic \times \Nc \rightarrow \R$, such that
\beqs
c_t &=& f_c(t,W_{. \wedge t}, I_{. \wedge t},N_{. \wedge t}), \;\; \; \mbox{ for } \P-a.e \;\; \omega, \;\;\; \mbox{for all }t \in \R_+\\
\enqs
\end{Lemma}
{\bf \noindent Proof.}
For the progressively measurable part one can see e.g. Theorem 2.10 in \cite{YongZhou}.
For $c$ predictable, notice that this is true if $c= X \mathbf{1}_{(t,s]}$, where $X$ is $\Fc^0_t$-measurable, and conclude with a monotone class argument.
\ep

\vspace{7mm}

{\bf \noindent Proof of Proposition A.1.} Let $V_i(x,y)$ be the right hand side of \reff{DPP2}.

\noindent {\it Step 1.}  $v_i(x,y) \leq V_i(x,y)$: Take $\Uc \in \Ac_i^w(x,y)$. Then
\beq
\E\left[ \int_0^\infty e^{-\rho t} U(c_t) dt \left| \Fc_\tau \right.  \right] = \int_0^\tau e^{-\rho t} U(c_t) dt + e^{-\rho \tau} \E\left[ \int_0^\infty e^{-\rho s} U(c_{\tau+s}) ds \left| \Fc_\tau \right.  \right]. \label{dppproof}
\enq
By Remark \ref{remF0}, w.l.o.g. we can assume that $c$ is $\F^0$-progressively measurable (resp. $\zeta$ $\F^0$-predictable).
For $\omega_0 \in \Omega$, define the shifted control $\tilde{\Uc}^{\omega_0} = (\Omega, \tilde{F}^\tau, \P_{\omega_0}, \tilde{\Fc}^\tau_t, \tilde{W}, \tilde{I}, \tilde{N}, \tilde{c}, \tilde{\zeta})$, where :
\begin{itemize}
	\item $\P_{\omega_0} = \P(.|\Fc_{\tau})(\omega_0)$
	\item $\tilde{W_t} = W_{\tau+t} - W_\tau$
	\item $\tilde{I_t} = I_{\tau+t}$
	\item $N'_t = N_{\tau+t} - N_\tau$
	\item $\tilde{F}^\tau$ is the augmentation of $\Fc$ by the $\P_{\omega_0}$-null sets, and $\tilde{\Fc_t^\tau}$ is the augmented filtration generated by $(\tilde{W}, \tilde{I}, \tilde{N})$.
	\item $\tilde{c_t} = c_{t+\tau}$, $\tilde{\zeta_t} = \zeta_{t+\tau}$
\end{itemize}
Then we can check that for almost all $\omega_0$, $\tilde{\Uc}^{\omega_0}$ satisfies the conditions of Definition \ref{defWeakControl} (with initial conditions $(I_{\tau}(\omega_0),X_{\tau}(\omega_0),Y_{\tau}(\omega_0))$) : 2. comes from the independence of $W$ and $(I,N)$ and the strong Markov property, and 4. is verified because for almost all $\omega_0$ $\Fc^0_{t+\tau} \subset \tilde{\Fc}^\tau_t$.

Moreover,  there is a modification $(X',Y')$ of $(X,Y)$ s.t. $(X'_{\tau+t},Y'_{\tau+t})$ is $\tilde{F}^\tau$-adapted, and a solution of \reff{dynX}-\reff{dynY} for $(\tilde{W}, \tilde{I}, \tilde{N})$. Hence $\tilde{\Uc}^{\omega_0} \in \Ac_{I_{\tau}(\omega_0)}^w(X_{\tau}(\omega_0),Y_{\tau}(\omega_0))$, and
\beqs
\E\left[ \int_0^\infty e^{-\rho s} U(c_{\tau+s}) ds \left| \Fc_\tau \right.  \right](\omega_0) = J(\tilde{\Uc}^{\omega_0}) \leq v_{I_{\tau}}(X_{\tau},Y_{\tau})(\omega_0).
\enqs
Hence taking the expectation over $\omega_0$ in \reff{dppproof},
\beqs
\E\left[ \int_0^\infty e^{-\rho t} U(c_t) dt \right] \leq \E\left[ \int_0^\tau e^{-\rho t} U(c_t) dt + e^{-\rho \tau}v_{I_{\tau}}(X_{\tau},Y_{\tau}) \right],
\enqs
and taking the supremum over $\Uc$, we obtain $v_i(x,y) \leq V_i(x,y)$.

\vspace{7mm}

\noindent {\it Step 2.}  $v_i(x,y) \geq V_i(x,y)$: Recall that in the proof of Proposition \ref{propv} we only needed the DPP to prove the continuity of $v_i$ up to the boundary. Hence we know a priori that $v_i$ is continuous on $Int(\R_+^2)$, and that the restriction of $v_i$ to the boundary is continuous.
One can then  find a countable sequence $(U_k)_{k \geq 0}$ s.t.
\begin{enumerate}
	\item [(i)] $(U_k)_k$ is a partition of $\R_+^2$,
	\item [(ii)] $\forall (x,y), (x',y') \in U_k, \forall i, \left| v_i(x,y) - v_i(x',y') \right| \leq \varepsilon$,
	\item [(iii)] $U_k$ contains its bottom-left corner $(x_k,y_k) = \left( \min_{(x,y) \in U_k} x, \min_{(x,y) \in U_k} y \right)$.
\end{enumerate}
Indeed, we can construct such a partition in the following way: $v_i$ is continuous on the boundary so we can partition each of the boundary lines into a countable number of segments verifying (ii) and (iii). Then in the interior we have first a partition in ``squared  rings" : $Int(\R_+^2) = \cup_{n \ge 1} K_n$, where $K_n = [1/(n+1),n+1]^2 \setminus [1/n,n]^2$. Since $v_i$ is continuous on the interior, we can partition each $K_n$ into a finite number of squares verifying (ii) and (iii). By taking the union of the line segments and the squares  for each $K_n$, we obtain  a sequence $(U_k)$ satisfying (i)-(iii).

Notice that (iii) implies the inclusion  $\Ac_i(x_k,y_k) \subset \Ac_i(x,y)$, for all $(x,y) \in U_k$.
For each $k$, take $\Uc^{i,k} = (\Omega^{i,k}, \Fc^{i,k}, \P^{i,k}, \F^{i,k}, W^{i,k}, I^{i,k}, N^{i,k}, c^{i,k}, \zeta^{i,k})$ $\varepsilon$-optimal for $(i,x_k,y_k)$, and $f_c^{i,k}, f_{\zeta}^{i,k}$ associated to $(c^{i,k},\zeta^{i,k})$ by Lemma \ref{lemCanonical}.
Then for each $(c,\zeta) \in \Ac_i(x,y)$, let us  define $\tilde{c}, \tilde{\zeta}$ by :
\beqs
\tilde{c}_t = \left\{\begin{array}{ll}c_t & \hbox{ when } t< \tau\\
 f_c^{i,k}(t - \tau, \tilde{W}(.\wedge(t - \tau)),\tilde{I}(.\wedge(t - \tau)),\tilde{N}(.\wedge(t - \tau))) & \hbox{ when } t\geq \tau, I_\tau=i, (X_{\tau},Y_{\tau}) \in U_k.
\end{array} \right.
\enqs
Then $\tilde{c}$ (resp. $\tilde{\zeta}$) is $\F$- progressively measurable (resp. predictable). Furthermore, for almost all $\omega_0$, with $i=I_\tau(\omega_0)$ and $(X_\tau,Y_\tau)(\omega_0) \in U_k$,
\beqs \Lc_{\P^{\omega_0}}(\tilde{W},\tilde{I},\tilde{N}, (\tilde{c}_{t+\tau}), (\tilde{\zeta}_{t+\tau})) = \Lc_{\P^{i,k}}(W^{i,k}, I^{i,k}, N^{i,k}, c^{i,k}, \zeta^{i,k}),
\enqs
and since $\Ac_i(x_k,y_k) \subset \Ac_{I_\tau(\omega_0)}(X_\tau(\omega_0),Y_\tau(\omega_0))$, this implies $X^{\tilde{c},\tilde{\zeta}}_t, Y^{\tilde{c},\tilde{\zeta}}_t \geq 0$ a.s., and $(\tilde{c},\tilde{\zeta}) \in \Ac_i(x,y)$. We also have
\beqs
\E\left[ \int_0^\infty e^{-\rho s} U(\tilde{c}_{\tau+s}) ds \left| \Fc_\tau \right.  \right](\omega_0) &=& \E^{i,k}\left[ \int_0^\infty e^{-\rho s} U(c^{i,k}_s) ds  \right] \\
 &\geq& v_i(x_k,y_k) - \varepsilon \\
 &\geq& v_{I_\tau}(X_\tau,Y_\tau)(\omega_0) - 2 \varepsilon.
\enqs
By  taking expectation in \reff{dppproof}, we have
\beqs
\E\left[ \int_0^\infty e^{-\rho t} U(\tilde{c}_t) dt\right] \geq \E\left[ \int_0^\tau e^{-\rho t} U(c_t) dt + e^{-\rho \tau} v_{I_\tau}(X_\tau,Y_\tau)  \right] - 2 \varepsilon.
\enqs
Finally,  by taking the supremum over $\Uc$,  and  letting $\varepsilon$ go to $0$, we obtain $v_i(x,y) \geq V_i(x,y)$.
\ep

\begin{Remark}
{\rm Actually the weak value function is equal to the value function defined in \reff{defVi} for any $(\Omega, \Fc, \P, \F, W, I, N)$ satisfying (1)-(3) in Definition \ref{defWeakControl}.
Indeed, given any $\Uc' = (\Omega', \Fc', \P', \F', W', I', N') \in \Ac_i^w(x,y)$, letting $f_{c'}$ and $f_{\zeta'}$ being associated to $c'$ and $\zeta'$ by Lemmas \ref{lemFtoF0} and \ref{lemCanonical}, and defining (almost surely) $c_t = f_{c'}(t,W,I,N)$,
$\zeta_t = f_{\zeta'}(t,W,I,N)$, by the same arguments as in the Proof of Proposition \ref{propDPPA}, $\Uc :=(\Omega, \Fc, \P, \F, W, I, N, c, \zeta) \in \Ac_i^w(x,y)$, and $J(\Uc)=J(\Uc')$.
Hence
\beqs
\sup_{\Uc' \in \Ac^w_i(x,y)} J(\Uc') = \sup_{(c,\zeta) \in \Ac_i(x,y)} \E\left[\int_0^{\infty} e^{-\rho s} U(c_s) ds \right].
\enqs
\hfill\qed
}
\end{Remark}

\bigskip

\section*{Appendix B: Viscosity characterization}

\renewcommand{\theLemma}{B.\arabic{Lemma}}
\renewcommand{\theDefinition}{B.\arabic{Definition}}
\renewcommand{\theProposition}{B.\arabic{Proposition}}
\renewcommand{\theTheorem}{B.\arabic{Theorem}}
\renewcommand{\theequation}{B.\arabic{equation}}
\renewcommand{\theRemark}{B.\arabic{Remark}}
\renewcommand{\thesection}{B.\arabic{section}}

\setcounter{section}{0}
\setcounter{equation}{0}
\setcounter{Lemma}{0}
\setcounter{Definition}{0}
\setcounter{Proposition}{0}
\setcounter{Theorem}{0}
\setcounter{Remark}{0}

We first prove the viscosity property of the value function to its dynamic programming system \reff{eqHJB}, written as:
\beqs
F_i(x,y,v_i(x,y),Dv_i(x,y),D^2v_i(x,y)) +  G_i(x,y,v) &=& 0, \; (x,y) \in (0,\infty)\times\R_+,
\enqs
for any $i$ $\in$ $\I_d$, where  $F_i$ is the local operator defined by:
\beqs
F_i(x,y,u,p,A) &=& \rho u - b_i y p_2 - \frac{1}{2} \sigma_i^2 y^2 a_{22} - \tilde U(p_1)
\enqs
for $(x,y)$ $\in$ $(0,\infty)\times\R_+$,  $u$ $\in$ $\R$, $p$ $=$ $(p_1 \; p_2)$ $\in$ $\R^2$, $A$ $=$
$\left(\begin{array}{cc}a_{11} & a_{12} \\a_{12} & a_{22}\end{array}\right)$ $\in$ $\Sc^2$ (the set of symmetric $2\times 2$ matrices),  and
$G_i$ is the nonlocal operator defined  by:
\beqs
G_i(x,y,w) &=&  - \sum_{j\neq i} q_{ij} \big[  w_j(x,y(1-\gamma_{ij})) - w_i(x,y)\big]
-  \lambda_i \big[ \hat w_i(x + y) - w_i(x,y) \big]
\enqs
for $w$ $=$ $(w_i)_{i\in\I_d}$ $d$-tuple of continuous functions on $\R_+^2$.

\begin{Proposition} \label{propVisc}
The value function $v$ $=$ $(v_i)_{i\in\I_d}$ is a viscosity solution of (E).
\end{Proposition}
{\bf Proof.} {\it Viscosity supersolution}:  Let  $(i,\bar x,\bar y)$ $\in$ $\I_d \times (0,\infty) \times \R_+$, $\varphi$ $=$ $(\varphi_i)_{i\in\I_d}$,  $C^2$ test functions s.t. $v_i(\bar x,\bar y)$ $=$ $\varphi_i(\bar x,\bar y)$, and $v$ $\geq$ $\varphi$.   Take some arbitrary  $e$ $\in$ $(-\bar y,\bar x)$, and $c$ $\in$ $\R_+$. Since $\bar x >0$,  there exists a strictly positive stopping time
$\tau$ $>$ $0$ a.s. such that the control process $(\bar\zeta,\bar c)$ defined by:
\beq \label{barc}
\bar\zeta_t \; = \; e 1_{t\leq\tau}, & & \bar c_t \; = \: c 1_{t\leq\tau},  \;\;\; t \geq 0,
\enq
with associated state process $(\bar X,\bar Y,I)$ starting from $(x,y,i)$ at time $0$, satisfies $\bar X_t$ $\geq$ $0$, $\bar Y_t$ $\geq$ $0$, for all $t$. Thus, $(\bar\zeta,\bar c)$ $\in$ $\Ac_i(x,y)$.
Let $\Vc$ be a compact neighbourhood of $(x,y,i)$ in $(0,\infty) \times \R_+\times\I_d$, and consider the sequence of stopping time:
$\theta_n$ $=$ $\theta\wedge h_n$, where $\theta$ $=$ $\inf \left\{ t \geq 0: (\bar X_t,\bar Y_t,I_t) \notin \Vc \right\}$, and $(h_n)$ is a strictly positive sequence converging to zero.
From the dynamic programming principle \reff{DPP}, and by applying It\^o's formula to $e^{-\rho t}\varphi(\bar X_t,\bar Y_t,I_t)$ between $0$ and $\theta_n$, we get:
\beqs
\varphi(\bar x,\bar y,i) \; = \; v(x,y,i) &\geq& \E\left[ \int_0^{\theta_n} e^{-\rho t} U(\bar c_t) dt + e^{-\rho \theta_n} v(\bar X_{\theta_n},\bar Y_{\theta_n},I_{\theta_n}) \right] \\
&\geq& \E\Big[ \int_0^{\theta_n} e^{-\rho t} U(\bar c_t) dt + e^{-\rho \theta_n} \varphi(\bar X_{\theta_n},\bar Y_{\theta_n},I_{\theta_n})\Big] \\
 &=& \varphi(\bar x,\bar y,i) + \E\Big[ \int_0^{\theta_n} e^{-\rho t} \Big(U(\bar c_t) -   \rho \varphi    -   \bar c_t  \Dx{\varphi}  \\
 & & \;\;\;\;\; \hspace{2cm}   + \;    b_{I_{t^-}} \bar Y_{t^-} \Dy{\varphi}   +  \frac{1}{2}\sigma_{I_{t^-}}^2 \bar Y_{t^-}^2 \Dyy{\varphi}  \\
& & \;\;\;\;\;\;\;\;\;   + \sum_{j\neq I_{t^-} } q_{_{I_{t^-}j}} [ \varphi(\bar X_{t^-},\bar Y_{t^-}(1-\gamma_{_{I_{t^-}j}}),j) - \varphi(\bar X_{t^-},\bar Y_{t^-},I_{t^-})]  \\
& &  \;\;\;\;\;\;\;\;\;   +  \lambda_{_{I_{t^-}}} \big[ \varphi(\bar X_{t^-} - \bar\zeta_t,\bar Y_{t^-}+\bar\zeta_t,I_{t^-}) - \varphi(\bar X_{t^-},\bar Y_{t^-},I_{t^-}) \big] \Big) dt \Big],
\enqs
and so
\beq
 \E\Big[ \frac{1}{h_n} \int_0^{\theta_n} e^{-\rho t} \Big(   \rho \varphi - U(\bar c_t)   +   \bar c_t  \Dx{\varphi} -     b_{I_{t^-}} \bar Y_{t^-} \Dy{\varphi}
 -  \frac{1}{2}\sigma_{I_{t^-}}^2 \bar Y_{t^-}^2 \Dyy{\varphi}  & & \nonumber \\
- \sum_{j\neq I_{t^-} } q_{_{I_{t^-}j}} [ \varphi(\bar X_{t^-},\bar Y_{t^-}(1-\gamma_{_{I_{t^-}j}}),j) - \varphi(\bar X_{t^-},\bar Y_{t^-},I_{t^-})]  & & \nonumber \\
- \lambda_{_{I_{t^-}}} \big[ \varphi(\bar X_{t^-} - \bar\zeta_t,\bar Y_{t^-}+\bar\zeta_t,I_{t^-}) - \varphi(\bar X_{t^-},\bar Y_{t^-},I_{t^-}) \big] \Big) dt \Big] & \geq & 0   \label{viscosur}
\enq
Now, we have almost surely for $n$ large enough, $\theta$ $\geq$ $h_n$, i.e. $\theta_n$ $=$ $h_n$,  so that by using also \reff{barc}
\beqs
& &  \frac{1}{h_n} \int_0^{\theta_n} e^{-\rho t} \Big(   \rho \varphi - U(\bar c_t)   +   \bar c_t  \Dx{\varphi} -     b_{I_{t^-}} \bar Y_{t^-} \Dy{\varphi}
 -  \frac{1}{2}\sigma_{I_{t^-}}^2 \bar Y_{t^-}^2 \Dyy{\varphi}  \nonumber \\
& & - \sum_{j\neq I_{t^-} } q_{_{I_{t^-}j}} [ \varphi(\bar X_{t^-},\bar Y_{t^-}(1-\gamma_{_{I_{t^-}j}}),j) - \varphi(\bar X_{t^-},\bar Y_{t^-},I_{t^-})]   \\
& & - \lambda_{_{I_{t^-}}} \big[ \varphi(\bar X_{t^-} - \bar\zeta_t,\bar Y_{t^-}+\bar\zeta_t,I_{t^-}) - \varphi(\bar X_{t^-},\bar Y_{t^-},I_{t^-}) \big] \Big) dt \Big] \\
& \longrightarrow & \rho \varphi_i(\bar x,\bar y) -U(c) + c \Dx{\varphi_i}(\bar x,\bar y) - b_i \bar y \Dy{\varphi_i}(\bar x,\bar y) - \frac{1}{2} \sigma_i^2 \bar y^2 \Dyy{\varphi_i}(\bar x,\bar y)  \\
& & - \sum_{j\neq i} q_{ij} [\varphi_j(\bar x,\bar y(1-\gamma_{ij}))-\varphi_i(\bar x,\bar y) ]  - \lambda_i [ \varphi_i(\bar x - e,\bar y + e) - \varphi_i(\bar x,\bar y) ], \;\;\; a.s.
\enqs
when $n$ goes to infinity. Moreover, since the integrand of the  Lebesgue integral term in \reff{viscosur} is bounded  for $t$ $\leq$ $\theta$,
 one can apply the dominated convergence theorem in \reff{viscosur}, which gives:
\beqs
\rho \varphi_i(\bar x,\bar y) - U(c) + c \Dx{\varphi_i}(\bar x,\bar y) - b_i \bar y \Dy{\varphi_i}(\bar x,\bar y) - \frac{1}{2} \sigma_i^2 \bar y^2 \Dyy{\varphi_i}(\bar x,\bar y) & &   \\
 - \sum_{j\neq i} q_{ij} [\varphi_j(\bar x,\bar y(1-\gamma_{ij}))-\varphi_i(\bar x,\bar y) ]  - \lambda_i [ \varphi_i(\bar x - e,\bar y + e) - \varphi_i(\bar x,\bar y) ] & \geq & 0.
\enqs
Since $c$ and $e$ are arbitrary, we obtain the required viscosity supersolution inequality by taking the supremum over $c$ $\in$ $\R_+$ and $e$ $\in$ $(-\bar y,\bar x)$.

  \vspace{3mm}

\noindent {\it Viscosity subsolution}: Let  $(\bar i,\bar x,\bar y)$ $\in$ $\I_d \times (0,\infty) \times \R_+$, $\varphi$ $=$ $(\varphi_i)_{i\in\I_d}$,  $C^2$ test functions s.t.
$v(\bar x,\bar y,\bar i)$ $=$ $\varphi(\bar x,\bar y,\bar i)$, and $v$ $\leq$ $\varphi$.  We can also assume w.l.o.g. that $v$ $<$ $\varphi$ outside $(\bar x,\bar y,\bar i)$.
We argue by contradiction by assuming that
\beqs
\rho \varphi_{\bar i}(\bar x,\bar y)   - b_{\bar i} \bar y \Dy{\varphi_{\bar i}}(\bar x,\bar y) - \frac{1}{2} \sigma_{\bar i}^2 \bar y^2 \Dyy{\varphi_{\bar i}}(\bar x,\bar y)
- \tilde U\Big( \Dx{\varphi_{\bar i}}(\bar x,\bar y) \Big) & &   \\
 - \sum_{j\neq \bar i} q_{\bar ij} [\varphi_j(\bar x,\bar y(1-\gamma_{\bar ij}))-\varphi_{\bar i}(\bar x,\bar y) ]  - \lambda_{\bar i} [ \hat\varphi_{\bar i}(\bar x +\bar y)
 - \varphi_{\bar i}(\bar x,\bar y) ] & >  & 0.
\enqs
By continuity of $\varphi$, and of its derivatives, there exist some compact neighbourhood $\bar\Vc$ of $(\bar x,\bar y,\bar i)$ in $(0,\infty) \times \R_+\times\I_d$, and $\eps$ $>$ $0$,
such that
\beq
\rho \varphi_i(x,y)  - b_i y \Dy{\varphi_i}(x,y) - \frac{1}{2} \sigma_i^2 y^2 \Dyy{\varphi_i}(x,y) - \tilde U\Big( \Dx{\varphi_{i}}(x,y) \Big) & &   \label{viscosous} \\
 - \sum_{j\neq i} q_{ij} [\varphi_j(x,y(1-\gamma_{ij}))-\varphi_i(x,y) ]  - \lambda_i [ \hat\varphi_i(x +  y) - \varphi_i(x,y) ] & \geq & \eps, \;\;\; \forall (x,y,i) \in  \bar\Vc.  \nonumber
\enq
Since $v$ $<$ $\varphi$ outside $(\bar x,\bar y,\bar i)$, there exists some $\delta$ $>$ $0$ s.t.  $v$ $<$ $\varphi - \delta$ outside of $\bar\Vc$. We can also assume that $\varepsilon \leq \delta \rho$.
By the DPP \reff{DPP},  there exists $(\zeta,c) \in \Ac_{\bar i}(\bar x,\bar y)$ s.t.
\beqs
v(\bar x,\bar y,\bar i) - \varepsilon\frac{1 - e^{-\rho}}{2 \rho} &\leq&
\E\left[ \int_0^{\theta \wedge 1} e^{-\rho t} U(c_t) dt + e^{-\rho (\theta \wedge 1)} v(X_{\theta\wedge 1},Y_{\theta\wedge 1},I_{\theta\wedge 1}) \right],
\enqs
where $(X,Y,I)$ is controlled by $(\zeta,c)$, and  we take  $\theta$ $=$  $\inf \left\{ t \geq 0:  (X_t,Y_t,I_t) \notin \bar\Vc \right\}$. We then get:
\beqs
& & \varphi(\bar x,\bar y,\bar i)  - \varepsilon\frac{1 - e^{-\rho}}{2 \rho} \\
&=& v(\bar x,\bar y,\bar i) - \varepsilon\frac{1 - e^{-\rho}}{2 \rho} \\
&\leq& \E\left[ \int_0^{\theta\wedge 1} e^{-\rho t} U(c_t) dt + e^{-\rho (\theta\wedge 1)} \varphi(X_{\theta\wedge 1},Y_{\theta\wedge 1},I_{\theta\wedge 1})
- e^{-\rho \theta} \delta \mathbf{1}_{\left\{\theta <1 \right\}} \right] \\
&=&  \varphi(\bar x,\bar y,\bar i)  +  \E\Big[ \int_0^{\theta \wedge 1} e^{-\rho t} \Big(U(c_t) -   \rho \varphi    -    c_t  \Dx{\varphi}  \\
 & & \;\;\;\;\; \hspace{2cm}   + \;    b_{I_{t^-}}  Y_{t^-} \Dy{\varphi}   +  \frac{1}{2}\sigma_{I_{t^-}}^2  Y_{t^-}^2 \Dyy{\varphi}  \\
& & \;\;\;\;\;\;\;\;\;   + \sum_{j\neq I_{t^-} } q_{_{I_{t^-}j}} [ \varphi( X_{t^-}, Y_{t^-}(1-\gamma_{_{I_{t^-}j}}),j) - \varphi(X_{t^-},Y_{t^-},I_{t^-})]  \\
& &  \;\;\;\;\;\;\;\;\;   +  \lambda_{_{I_{t^-}}} \big[ \varphi(X_{t^-} - \zeta_t,Y_{t^-}+\zeta_t,I_{t^-}) - \varphi(X_{t^-},Y_{t^-},I_{t^-}) \big] \Big) dt - e^{-\rho \theta} \delta \mathbf{1}_{\left\{\theta <1 \right\}}   \Big] \\
&\leq &  \varphi(\bar x,\bar y,\bar i)  + \E\left[\int_0^{\theta\wedge 1} - \varepsilon e^{-\rho t}dt - e^{-\rho \theta} \delta \mathbf{1}_{\left\{\theta<1 \right\}} \right]
\enqs
where we applied  It\^o's  formula in the second equality, and used \reff{viscosous} in the last inequality.  This means that
\beqs
- \varepsilon\frac{1 - e^{-\rho}}{2 \rho} &\leq& \E\left[\int_0^{\theta\wedge 1} - \varepsilon e^{-\rho t}dt - e^{-\rho \theta} \delta \mathbf{1}_{\left\{\theta<1 \right\}} \right] \\
&=& \E\left[- \frac{\varepsilon}{\rho} +  \frac{\varepsilon}{\rho}e^{-\rho(\theta \wedge 1)} - e^{-\rho \theta} \delta \mathbf{1}_{\left\{\theta<1 \right\}} \right] \; \leq \;  - \frac{\varepsilon}{\rho}(1-e^{-\rho}),
\enqs
since $\eps/\rho$ $\leq$ $\delta$, and we get the required contradiction.
\ep

\vspace{7mm}

Let us now  prove comparison principle for our dynamic programming system. As usual, it is convenient to formulate an equivalent definition for
viscosity solutions to \reff{eqHJB} in terms of semi-jets.  We shall  use the notation $X = (x,y)$ for $\R_+ \times \R_+$-valued vectors.
Given $w$ $=$ $(w_i)_{i\in\I_d}$ a d-tuple of continuous functions on $\R_+^2$,  the second-order {\it superjet} of $w_i$  at
$X$ $\in$ $\R_+^2$ is defined by:
\beqs
\Pc^{2,+}w_i(X) &=&  \Big\{(p,A) \in  \R^2 \times \Sc^2 \mbox{   s.t.  } w_i(X') \leq w_i(X) + \left\langle p, X' -X\right\rangle  \\
                         &&  \;\;\;\;\;  + \frac{1}{2}\left\langle A(X'-X), X' -X\right\rangle\ + o\left(\left|X'-X\right|^2 \right) \mbox{ as }X' \rightarrow X \Big\},
\enqs
and its closure $\overline{\Pc}^{2,+}w_i(X)$ as the set  of  elements $(p,A)$ $\in$ $\R^2\times\Sc^2$ for which there exists a sequence
$(X_m,p_m,A_m)_m$ of $\R_+^2\times\Pc^{2,+}w_i(X_m)$ satisfying  $(X_m,p_m,A_m)$ $\rightarrow$ $(X,p,A)$.
We also define the second-order
\emph{subjet} $\Pc^{2,-}w_i(X)$ $=$ $-\Pc^{2,+}(-w_i)(X)$, and $\overline{\Pc}^{2,-}w_i(X)$ $=$ $-\overline{\Pc}^{2,+}(-w_i)(X)$.
By standard arguments (see e.g. \cite{ari08} for  equations with nonlocal terms), one has an equivalent definition of viscosity solutions in terms of semijets:

\vspace{2mm}

\noindent A $d$-tuple $w$ $=$ $(w_i)_{i\in\I_d}$ of continuous functions on $\R_+^2$ is a viscosity supersolution (resp. subsolution) of \reff{eqHJB}  if and only if  for all  $(i,x,y) \in \I_d \times (0,\infty) \times \R_+$,  and all $(p,A)$ $\in$ $\overline{\Pc}^{2,-}w_i(x,y)$ (resp.  $\overline{\Pc}^{2,+}w_i(x,y)$):
\beqs
F_i(x,y,w_i(x,y),p,A) +  G_i(x,y,w) & \geq&  0,  \;\;\; (resp. \; \leq \; 0).
\enqs

 \vspace{1mm}

 We then prove the following comparison theorem.

\begin{Theorem} \label{thmComparison}
Let $V$ $=$ $(V_i)_{i\in\I_d}$  (resp. $W$ $=$ $(W_i)_{i\in\I_d}$) be a viscosity subsolution (resp. supersolution) of \reff{eqHJB}, satisfying the growth condition \reff{growthvi}, and the boundary conditions
\beq
V_i(0,0) &\leq& 0 \label{ineqBndry0} \\
V_i(0,y) &\leq& \E_i\left[ \hat V_{_{I^i_{\tau_1}}}\big(y\frac{S_{\tau_1}}{S_0}\big) \right], \;\;\;  \forall y > 0,   \label{ineqBndry1}
\enq
(resp. $\geq$ for $W$).  Then $V$ $\leq$ $W$.
\end{Theorem}
{\bf Proof.} {\it Step 1}:  Take $p' >p$ such that $k(p') < \rho$, and define $\psi_i(x,y) = (x+y)^{p'}$, $i$ $\in$ $\I_d$.
Let us check that  $W^n$ $=$ $W + \frac{1}{n} \psi$ is still a supersolution of (E).  Notice that  $\Pc^{2,-}W_i^n$ $=$
$\Pc^{2,-}W_i + \frac{1}{n}(D\psi_i, D^2 \psi_i)$, and we have for all  $(p,A) \in \Pc^{2,-}W_i(x,y)$:
\beq
&&F_i\big(x,y,W^n_i(x,y), p + \frac{1}{n} D\psi_i, A + \frac{1}{n} D^2 \psi_i\big) +  G_i(x,y,W^n) \nonumber \\
&=& F_i\big(x,y,W_i(x,y), p, A) +  G_i(x,y,W) \nonumber \\
& & \;  + \; \frac{1}{n}(x+y)^{p'} \Big( \rho - p'b_i \frac{y}{x+y} + p'(1-p') \frac{\sigma_i^2}{2} \left(\frac{y}{x+y}\right)^2 - \sum_{j \neq i} q_{ij}((1 - \frac{y}{x+y} \gamma_{ij})^{p'} - 1)  \Big) \nonumber \\
&& \; + \;  \tilde{U}(p_1) - \tilde{U}\big(p_1 + \frac{1}{n}p' x^{p'-1}\big) \label{calcWn} \\
& \geq & 0.  \nonumber
\enq
Indeed, the three lines in the r.h.s. of  \reff{calcWn} are nonnegative: the first one since $W$ is a supersolution, the second one by $k(p')<\rho$, and the last one since $\tilde{U}$ is nonincreasing.

\vspace{1mm}

Moreover, by the growth condition \reff{growthvi} on $V$ and $W$, we have:
\beq
\lim_{r \rightarrow \infty} \max_{i \in \I_d} (\hat V_i - \hat W_i^n)(r) &=& - \infty. \label{limVW}
\enq
In the next step, our aim is to  show that for all $n$ $\geq$ $1$, $V$ $\leq$ $W^n$, which would imply that $V$ $\leq$ $W$.  We shall argue by contradiction.

\noindent {\it Step 2}:  Assume that there exists  some $n \geq 1$ s.t.
\beqs
M := \sup_{i\in\I_d, (x,y)\in\R_+^2} (V_i - W_i^n)(x,y) >0.
\enqs
By \reff{limVW}, there exists $i \in \I_d$, some compact subset $\Cc$ of $\R_+^2$, and $\overline{X}=(\overline{x},\overline{y}) \in \Cc$ such that
\beq \label{Mbar}
M &=& \max_{\Cc} (V_i - W^n_i) \; = \;  (V_i-W^n_i)(\overline{x},\overline{y}).
\enq
Note that by \reff{ineqBndry0}, $(\overline{x},\overline{y}) \neq (0,0)$.  We then have two possible cases:

\vspace{1mm}

\noindent  $\bullet$ Case 1 : $\overline{x} = 0$.  Notice that the boundary condition \reff{ineqBndry1} implies the viscosity subsolution property
for $V_i$ also at $\bar X$ $=$ $(0,\bar y)$:
\beqs \label{viscos0}
F_i(\bar X,V_i(\bar X),p,A) +  G_i(\bar X,V) & \leq&  0,  \;\;\; \forall (p,A) \in  \overline{\Pc}^{2,+}V_i(\bar X)
\enqs
However  the viscosity supersolution property fot $W^n$ does not hold at $(0,\bar y)$. Let $(X_k)_k=(x_k,y_k)_k$ be a sequence converging to $\overline{X}$, with $x_k > 0$, and $\eps_k:= \left|X_k - \overline{X}\right|$. We then consider the function

\beqs
\Phi_{k}(X,X') &=& V_i(X) - W^n_i(X') - \psi_k(X,X'), \\
\psi_k(x,y,x',y') &=& x^4 + (y - \overline{y})^4 + \frac{\left|X - X'\right|^2}{2 \eps_k} + \left(\frac{x'}{x_k}-1 \right)^3_-
\enqs
Since $\Phi_k$ is continuous, there exists $(\widehat X_k,\widehat X'_k)$ $\in$ $\Cc^2$ s.t.
\beqs
M_k := \sup_{\Cc^2} \Phi_k = \Phi_k(\widehat X_k,\widehat X'_k),
\enqs
and a subsequence, still denoted $(\widehat X_k,\widehat X'_k)$, converging to some $(\widehat X,\widehat X')$ as $k$ goes to $\infty$.
By writing that $\Phi_k(\overline{X}, X_k) \leq \Phi_k(\widehat X_k,\widehat X'_k)$, we have :
\beq
& & V_i(\overline{X}) - W^n_i(X_k) - \frac{\left|\overline{X} - X_k\right|}{2} \label{ineq1} \\
&\leq& V_i(\widehat X_k) - W^n_i(\widehat X'_k) - (\hat x_k^4 + (\hat y_k - \overline{y})^4) - R_k \label{ineq2} \\
&\leq& V_i(\widehat X_k) - W^n_i(\widehat X'_k) - (\hat x_k^4 + (\hat y_k - \overline{y})^4) \label{ineq3},
\enq
where we set
\beqs
R_k &=& \frac{\left|\widehat X_k - \widehat X'_k\right|^2}{2 \eps_k} + \left(\frac{\hat{x}'_k}{x_k}-1 \right)^3_-
\enqs
Since $V_i$ and $W^n_i$ are bounded on $\Cc$, we deduce by inequality \reff{ineq2} the boundedness of the sequence $(R_k)_{k\geq 0}$, which implies $\widehat{X} = \widehat{X'}$. Then by sending $k$ to infinity in \reff{ineq1} and \reff{ineq3}, with the continuity of $V_i$ and $W^n_i$, we obtain $M = V_i(\overline{X}) - W^n_i(\overline{X}) \leq  V_i(\widehat X) - W^n_i(\widehat X) - (\hat x_k^4 + (\hat{y}_k - \overline{y})^4)$, and by definition of $M$ this shows
\beq
\widehat X = \widehat X'  =\overline{X} \label{convXk}
\enq
Sending again $k$ to infinity in \reff{ineq1}-\reff{ineq2}-\reff{ineq3}, we obtain $M \leq M - \limsup_k R_k \leq M$, and so
\beq
\frac{\left|\widehat X_k - \widehat X'_k\right|^2}{2 \eps_k} + \left(\frac{\widehat{x}'_k}{x_k}-1 \right)^3_- &\rightarrow& 0, \label{convRk}
\enq
as $k$ goes to infinity. In particular for $k$ large enough $\hat{x}'_k \geq \frac{x_k}{2} >0$. We can  then apply Ishii's lemma (see Theorem 3.2 in \cite{craishlio92}) to obtain $A,A' \in \Sc^2$ s.t.
\beq
\left(p, A\right) \in \overline{\Pc}^{2,+}V_i(\widehat{X}_k), \;\;\;\; \left(p', A'\right) \in \overline{\Pc}^{2,-}W^n_i(\widehat{X}'_k) \label{ABjet} \\
\left(\begin{array}{cc} A&0\\0&-A'\end{array}\right) \leq  D + \eps_k D^2 \label{ineqMat},
\enq
where
\beqs
p=D_X \psi_k(\widehat{X}_k,\widehat{X'}_k), \;\;\;\; p'=D_{X'} \psi_k(\widehat{X}_k,\widehat{X}'_k), \;
D = D^2_{X,X'}\psi_k(\widehat{X}_k,\widehat{X'}_k).
\enqs
Now, we write
\beq
\rho M \;  \leq \;  \rho M_k &\leq& \rho(V_i(\hat{X}_k) - W^n_i(\widehat{X'}_k)) \nonumber \\
&=& F_i\big(\widehat{X}_k,V_i(\widehat{X}_k),p,A\big) - F_i\big(\widehat{X}_k,W^n_i(\widehat{X'}_k),p,A\big) \nonumber \\
&=& F_i\big(\widehat{X}_k,V_i(\widehat{X}_k),p,A\big) + G_i(\widehat{X}_k,V)  \label{FF}  \\
 && \; - \; F_i\big(\widehat{X}'_k,W^n_i(\widehat{X}'_k),p',A'\big) -  G_i(\widehat{X}'_k,W^n) \nonumber \\
 && \; + \;  G_i(\widehat{X}'_k,W^n) -    G_i(\widehat{X}_k,V) \nonumber \\
 && \; + \;  F_i\big(\widehat{X}'_k,W^n_i(\widehat{X}'_k),p',A'\big) - F_i\big(\widehat{X}_k,W^n_i(\widehat{X}'_k),p,A\big) \nonumber
\enq
From the viscosity subsolution property  for $V$ at $\hat X_k$, and the viscosity supersolution property for $W^n$ at $\hat X_k'$,
the first two lines  in  the r.h.s. of \reff{FF} are nonpositive.  For the third line, by sending $k$ to infinity, we have:
\beqs
& & G_i(\widehat{X}'_k,W^n) -    G_i(\widehat{X}_k,V) \\
&\rightarrow & G_i(\overline{X},W^n) - G_i(\overline{X},V)   \\
&=&  \sum_{j\neq i} q_{ij} \Big[  (V_j-W^n_j) \Big(\overline{x},\overline{y}(1- \gamma_{ij})\Big) - (V_i-W^n_i)(\overline{x},\overline{y})  \Big]  \\
& & + \lambda_i \Big[ \big(\hat V_i  - \hat W^n_i\big)(\overline{x}+\overline{y}) -   (V_i-W^n_i)(\overline{x},\overline{y}) \Big] \\
& \leq & 0
\enqs
by \reff{Mbar}.  For the fourth line of \reff{FF}, we have
\beqs
& & F_i\big(\widehat{X}'_k,W^n_i(\widehat{X}'_k),p',A'\big) - F_i\big(\widehat{X}_k,W^n_i(\widehat{X}'_k),p,A\big) \\
&=&  b_i (\hat{y}_k p_2 - \hat{y}'_k p'_2) + \tilde{U}(p_1) - \tilde{U}(p'_1) + \frac{\sigma_i^2}{2}\left(\hat{y}_k^2 a_{22}
- (\widehat{y}'_k)^2 a'_{22}\right)
\enqs
Now
\beqs
 \hat{y}_k p_2 - \hat{y}'_k p'_2 &=& \widehat{y}_k \left( 4(\hat{y}_k - \overline{y})^3 + \frac{\hat{y}_k - \hat{y}'_k}{\eps_k} \right) - \hat{y}'_k \left(\frac{\hat{y}_k - \hat{y}'_k}{\eps_k} \right) \\
 &\leq& 4\hat{y}_k (\hat{y}_k - \overline{y})^3 + \frac{\left|\widehat{X}_k -\hat x'_k\right|^2}{\eps_k} \\
 &\rightarrow& 0,  \mbox{ as } \;  k \rightarrow \infty,
\enqs
by \reff{convXk} and \reff{convRk}. Moreover,
\beqs
\tilde{U}(p_1) - \tilde{U}(p'_1) &=& \tilde{U}\left(\frac{\hat{x}_k -\hat x'_k}{\eps_k} + 4\hat x_k^3 \right) - \tilde{U}\left(\frac{\hat{x}_k -\hat x'_k}{\eps_k} - \frac{3}{x_k}\left(\frac{\hat{x}'_k}{x_k}-1 \right)^2_-\right) \\
&\leq& 0,
\enqs
since $\tilde{U}$ is nonincreasing. Finally,
\beqs
\hat{y}_k^2 a_{22} - (\hat{y}'_k)^2 a'_{22} &=& \left(\begin{array}{cccc}0&\hat{y}_k&0&\hat{y}'_k\end{array}\right) \left(\begin{array}{cc} A&0\\0&-A'\end{array}\right)\left(\begin{array}{c}0\\ \hat{y}_k\\0 \\
\hat{y}'_k\end{array}\right) \\
&\leq& \left(\begin{array}{cccc}0&\hat{y}_k&0&\hat{y}'_k\end{array}\right) \left(D + \eps_k D^2\right)\left(\begin{array}{c}0\\ \hat{y}_k\\0 \\ \hat{y}'_k\end{array}\right)
\enqs
by \reff{ineqMat}. Since
\beqs
D^2 \psi_k(x,y,x',y') = \left(\begin{array}{cccc}12 x^2&0&-\frac{1}{\eps_k}&0\\0&12(y-\overline{y})^2 + \frac{1}{\eps_k}&0&- \frac{1}{\eps_k}\\-\frac{1}{\eps_k}&0&\frac{1}{\eps_k}+\frac{6}{x_k^2}\left(\frac{x'}{x_k}-1 \right)_-&0\\0&- \frac{1}{\eps_k}&0&-\frac{1}{\eps_k}  \end{array}\right),
\enqs
a direct calculation gives
\beqs
\left(\begin{array}{cccc}0&\hat{y}_k&0&\hat{y}'_k\end{array}\right) \left(D + \eps_k D^2\right)\left(\begin{array}{c}0\\ \hat{y}_k\\0 \\ \hat{y}'_k\end{array}\right)
&=& \frac{3}{\eps_k}(\hat{y}_k-\hat{y}'_k)^2  - 12 (\hat{y}_k - \overline{y})^2 \hat{y}_k \hat{y}'_k  \\
&& + \left(36(\hat{y}_k - \overline{y})^2 + \eps_k\left(12(\hat{y}_k - \overline{y})^2\right)\right) \hat{y}_k^2\\
&\rightarrow& 0, \;\;\; \mbox{  as  } k \rightarrow \infty,
\enqs
where we used again  \reff{convXk} and \reff{convRk}, and the boundedness of $(\widehat{y}_k,\widehat{y'}_k)$.

\vspace{3mm}

Finally by letting $k$ go to infinity in \reff{FF} we  obtain $\rho M \leq 0$, which is the required  contradiction.

\vspace{5mm}

\noindent $\bullet$ Case 2 : $\overline{x} > 0$. This is the easier case, and we can obtain a contradiction similarly as in the first case, by considering for instance the function
\beqs
\Phi_{k}(X,X') &=& V_i(X) - W^n_i(X') -  (x-\overline{x})^4 - (y - \overline{y})^4 - k \frac{\left|X - X'\right|^2}{2}.
\enqs
\ep

\bigskip

\begin{small}

\end{small}

 \end{document}